%% file: main.tex
\documentclass[fleqn,usenatbib]{mnras}

\usepackage{newtxtext,newtxmath}

\usepackage[T1]{fontenc}

\DeclareRobustCommand{\VAN}[3]{#2}
\let\VANthebibliography\thebibliography
\def\thebibliography{\DeclareRobustCommand{\VAN}[3]{##3}\VANthebibliography}

\usepackage{graphicx}	
\usepackage{amsmath}	

\newcommand{\beagle}{\mbox{\sc{Beagle}}}
\newcommand{\bluetides}{\mbox{\sc{BlueTides}}}
\newcommand{\eagle}{\mbox{\sc{Eagle}}}
\newcommand{\ceagle}{\mbox{\sc{C-Eagle}}}
\newcommand{\cigale}{\mbox{\sc{Cigale}}}
\newcommand{\flares}{\mbox{\sc Flares}}
\newcommand{\flare}{\mbox{\sc Flare}}
\newcommand{\firstlight}{\mbox{\sc FirstLight}}
\newcommand{\subfind}{\mbox{\sc Subfind}}
\newcommand{\cloudy}{\mbox{\sc cloudy}}

\newcommand{\nion}{$\dot{N}_{\rm ion}$}
\newcommand{\nionintr}{$\dot{N}_{\rm ion, intr}$}

\newcommand{\nionesc}{$\dot{N}_{\rm ion, esc}$}
\newcommand{\nionM}{$\dot{N}_{\rm ion, intr}/M_\star$}
\newcommand{\xiion}{$\xi_{\rm ion}$}
\newcommand{\xidust}{$\xi_{\rm ion, dust}$}
\newcommand{\xistellar}{$\xi_{\rm ion, stellar}$}

\newcommand{\oiii}{[\ion{O}{III}]}
\newcommand{\oiiihb}{[\ion{O}{III}]$+\rm{H}\beta$}
\newcommand{\lya}{Ly$\alpha$}
\newcommand{\ha}{H$\alpha$}
\newcommand{\hb}{H$\beta$}
\newcommand{\hg}{H$\gamma$}
\newcommand{\fesc}{$f_{\rm esc}$}

\title[FLARES XIII: LyC emission of high-z galaxies]{First Light And Reionisation Epoch Simulations (FLARES) XIII: The Lyman-continuum emission of high-redshift galaxies}

\author[Seeyave et al.]{Louise T. C. Seeyave$^{1}$\thanks{E-mail: L.Seeyave@sussex.ac.uk},
Stephen M. Wilkins$^{1,2}$, 
Jussi K. Kuusisto$^{1}$, 
Christopher C. Lovell$^{3,1}$, 
\newauthor
Dimitrios Irodotou$^{4}$, 
Charlotte Simmonds$^{5,6}$, 
Aswin P. Vijayan$^{7,8}$, 
Peter A. Thomas$^{1}$, 
William J. Roper$^{1}$, 
\newauthor
Conor M. Byrne$^{9}$, 
Gareth T. Jones$^{9}$, 
Jack C. Turner$^{1}$, 
Christopher J. Conselice$^{10}$
\\
$^{1}$Astronomy Centre, University of Sussex, Falmer, Brighton BN1 9QH, UK\\
$^{2}$Institute of Space Sciences and Astronomy, University of Malta, Msida MSD 2080, Malta\\
$^{3}$Institute of Cosmology and Gravitation, University of Portsmouth, Burnaby Road, Portsmouth, PO1 3FX, UK\\
$^{4}$Department of Physics, University of Helsinki, Gustaf Hällströmin katu 2, FI-00014, Helsinki, Finland\\
$^{5}$The Kavli Institute for Cosmology (KICC), University of Cambridge, Madingley Road, Cambridge, CB3 0HA\\
$^{6}$Cavendish Laboratory, University of Cambridge, 19 JJ Thomson Avenue, Cambridge, CB3 0HE, UK\\
$^{7}$Cosmic Dawn Center (DAWN)\\
$^{8}$DTU-Space, Technical University of Denmark, Elektrovej 327, DK-2800 Kgs. Lyngby, Denmark\\
$^{9}$Department of Physics, University of Warwick, Gibbet Hill Road, Coventry, CV4 7AL, UK\\
$^{10}$Jodrell Bank Centre for Astrophysics, Department of Physics and Astronomy, University of Manchester, Oxford Road, Manchester M13 9PL, UK
}

\date{Accepted XXX. Received YYY; in original form ZZZ}

\pubyear{2023}

\begin{document}
\label{firstpage}
\pagerange{\pageref{firstpage}--\pageref{lastpage}}
\maketitle

\begin{abstract} 
The history of reionisation is highly dependent on the ionising properties of high-redshift galaxies. It is therefore important to have a solid understanding of how the ionising properties of galaxies are linked to physical and observable quantities. In this paper, we use the First Light and Reionisation Epoch Simulations (\flares) to study the Lyman-continuum (LyC, i.e. hydrogen-ionising) emission of massive ($M_*>10^8\,\mathrm{M_\odot}$) galaxies at redshifts $z=5-10$. We find that the specific ionising emissivity (i.e. intrinsic ionising emissivity per unit stellar mass) decreases as stellar mass increases, due to the combined effects of increasing age and metallicity. \flares\ predicts a median ionising photon production efficiency (i.e. intrinsic ionising emissivity per unit intrinsic far-UV luminosity) of $\log_{10}(\xi_{\rm ion}\rm{/erg^{-1}Hz})=25.40^{+0.16}_{-0.17}$, with values spanning the range $\log_{10}(\xi_{\rm ion}\rm{/erg^{-1}Hz})=25-25.75$. This is within the range of many observational estimates, but below some of the extremes observed. We compare the production efficiency with observable properties, and find a weak negative correlation with the UV-continuum slope, and a positive correlation with the \oiii\ equivalent width. We also consider the dust-attenuated production efficiency (i.e. intrinsic ionising emissivity per unit dust-attenuated far-UV luminosity), and find a median of $\log_{10}(\xi_{\rm ion}\rm{/erg^{-1}Hz})\sim25.5$. Within our sample of $M_*>10^8\,\mathrm{M_\odot}$ galaxies, it is the stellar populations in low mass galaxies that contribute the most to the total ionising emissivity. Active galactic nuclei (AGN) emission accounts for $10-20$\% of the total emissivity at a given redshift, and extends the LyC luminosity function by $\sim0.5$ dex. 
\end{abstract} 



\input{sections/1.intro}
\input{sections/2.theory}
\input{sections/3.sims}
\input{sections/4.ippe_mstar}
\input{sections/5.ippe_fuv}

\input{sections/6.emissivity}
\input{sections/conclusion}

\section*{Acknowledgements}
We wish to thank the anonymous referee for comments and suggestions that improved the paper. We thank Aaron Yung for sharing data from his models, and Ryan Endsley, Michael Topping, and Andreas Faisst for sharing their observational data in electronic format. We thank the \eagle\ team for their efforts in developing the \eagle\ simulation code, as well as Scott Kay and Adrian Jenkins for their invaluable help getting up and running with the \eagle\ resimulation code. 

This work used the DiRAC@Durham facility managed by the Institute for Computational Cosmology on behalf of the STFC DiRAC HPC Facility (www.dirac.ac.uk). The equipment was funded by BEIS capital funding via STFC capital grants ST/K00042X/1, ST/P002293/1, ST/R002371/1 and ST/S002502/1, Durham University and STFC operations grant ST/R000832/1. DiRAC is part of the National e-Infrastructure. We also wish to acknowledge the following open source software packages used in the analysis: \textsc{Scipy} \cite[][]{2020SciPy-NMeth}, \textsc{Astropy} \cite[][]{Astropy:2022}, \textsc{Matplotlib} \cite[][]{Hunter:2007}. 

LTCS is supported by an STFC studentship. SMW, PAT, and WJR thank STFC for support through ST/X001040/1. APV acknowledges support from the Carlsberg Foundation (grant no CF20-0534). PAT acknowledges support from the Science and Technology Facilities Council (grant number ST/P000525/1). DI acknowledges support by the European Research Council via ERC Consolidator Grant KETJU (no. 818930) and the CSC – IT Center for Science, Finland. CCL acknowledges support from a Dennis Sciama fellowship funded by the University of Portsmouth for the Institute of Cosmology and Gravitation. The Cosmic Dawn Center (DAWN) is funded by the Danish National Research Foundation under grant No. 140.

We list here the roles and contributions of the authors according to the Contributor Roles Taxonomy (CRediT)\footnote{\url{https://credit.niso.org/}}.
\textbf{Louise T. C. Seeyave, Stephen M. Wilkins}: Conceptualization, Data curation, Methodology, Investigation, Formal Analysis, Visualization, Writing - original draft.
\textbf{Christopher C. Lovell,  Aswin P. Vijayan}: Data curation, Methodology, Writing - review \& editing.
\textbf{Jussi K. Kuusisto}: Data curation.
\textbf{Conor M. Byrne, Christopher J. Conselice, Dimitrios Irodotou, Gareth T. Jones, William J. Roper, Charlotte Simmonds, Peter A. Thomas, Jack C. Turner}: Writing - review \& editing.


\section*{Data Availability}

Some of the \flares\ data published this paper is publicly available at \href{https://github.com/louiseseeyave/FLARES_LyC/}{https://github.com/louiseseeyave/flares\_lyc}. Data from the wider \flares\ project is available at \href{https://flaresimulations.github.io/#data}{https://flaresimulations.github.io/\#data}.



\bibliographystyle{mnras}
\bibliography{cite, cite_packages}


\appendix

\input{sections/appendix}


\bsp	
\label{lastpage}
\end{document}

%% file: sections/1.intro.tex
\section{Introduction}\label{sec:intro}

The Epoch of Reionisation (EoR) is the period of cosmic history in which hydrogen in the intergalactic medium (IGM) transitioned from a neutral to ionised state. Understanding how this process occurred is one of the key goals of modern extragalactic astrophysics. In the prevailing model, reionisation is driven by ionising radiation from stars and active galactic nuclei (AGN) \citep{Dayal_2018, Robertson_2022}, and is complete in most regions by $z=5-6$ \citep{Fan_2006, McGreer_2015, Eilers_2018, Yang_2020, Choudhury_2021, Bosman_2022}. Various theoretical and observational studies have shown that stars are likely the dominant source of ionising photons \citep{Duncan_2015, Onoue_2017, Dayal_2020, Yung_JWST_V, Yeh_2023}. In addition, many models suggest a greater overall contribution from low-mass galaxies ($M_\star<10^9\,\rm{M_\odot}$), though uncertainties remain regarding the exact makeup of the ionising photon budget and how it changes with redshift \citep{Finkelstein_2019, Lewis_2020, Yung_JWST_IV, Bera_2022, Mutch_2023}.

An important parameter that describes the ionising output of a source is its escaping ionising emissivity \nionesc, defined as the rate at which escaping ionising photons are produced. Most ionising photons produced inside a galaxy are reprocessed by gas and dust in the interstellar medium (ISM). The escape fraction \fesc\ is the fraction of ionising photons that manage to escape the galactic environment. It follows that \nionesc\ is obtained by combining \fesc\ with the intrinsic ionising emissivity \nionintr, the rate at which all ionising photons are produced, including those that end up reprocessed:
\begin{equation}
    \dot{N}_{\rm ion,esc} = f_{\rm esc} \times \dot{N}_{\rm ion,intr}. \label{eq:nionesc}
\end{equation}
Note that \nionintr\ is the integral over the stellar spectral energy distribution (SED) of a galaxy above the Lyman limit (912\AA):
\begin{equation}
    \dot{N}_{\rm ion,intr}=\int_{\nu_{912}}^\infty L_\nu(h\nu)^{-1}\rm{d}\nu.
    \label{eq:nionintr}
\end{equation}
\nionintr\ is often quantified by the ionising photon production efficiency, \xiion. Depending on the context, \xiion\ can be defined as a normalisation by stellar mass $M_\star$, or more commonly amongst observers, the intrinsic far-UV luminosity $L_{\rm FUV}$ (measured at rest-frame 1500\AA). In this paper, we use the latter definition:
\begin{equation}
    \xi_{\rm ion}=\frac{\dot{N}_{\rm ion,intr}}{L_{\rm FUV}}.
    \label{eq:xiion}
\end{equation}
We refer to the former definition, i.e. $\dot{N}_{\rm ion,intr}/M_\star$, as the specific ionising emissivity. 

The ionising photon production efficiency has been the target of a number of observational studies in recent years, and the rapidly expanding availability of observations from the James Webb Space Telescope (JWST) is now enabling more robust constraints on this key parameter at redshifts relevant to reionisation. Numbers derived are generally in the range $\log_{10}(\xi_{\rm ion}\rm{/\,erg^{-1}\,Hz})=25-26$ \citep[e.g.][]{Shivaei_2018, Emami_2020, Castellano_2022}, although some studies have measured larger values, with $\log_{10}(\xi_{\rm ion}\rm{/\,erg^{-1}Hz})>26$ \citep{Endsley_2021, Ning_2022, Simmonds_2023}. The production efficiency is often inferred from stellar population synthesis (SPS) models through SED fitting \citep[e.g.][]{Castellano_2022, Endsley_2022, Tang_2023}, or using emission line fluxes, typically the Balmer recombination lines \citep[e.g.][]{Nakajima_2016, Matthee_2022, Fujimoto_2023}. In the former scenario, the intrinsic ionising emissivity $\dot{N}_{\rm ion, intr}$ is obtained from the fitted galaxy SED model, following Equation \ref{eq:nionintr}. In the latter scenario, $\dot{N}_{\rm ion, intr}$ can be estimated from the flux of the Balmer lines, using a conversion factor supplied by stellar evolution models \citep[e.g.][]{Leitherer_1995, Schaerer_2003}.

Theoretical studies have also played a part in developing our understanding of the production efficiency of distant galaxies. \cite{Wilkins_2016} modelled the ionising photon production efficiency of galaxies in the \bluetides\ simulations, predicting values of $\log_{10}(\xi_{\rm ion}\rm{/\,erg^{-1} Hz})\approx25.1-25.5$ for a range of stellar population synthesis (SPS) models. A similar spread of values was obtained by \cite{Ceverino_2019}, who modelled the SEDs of galaxies in the \firstlight\ simulations, and by \cite{Yung_2020}, who used a semi-analytic modelling approach. Both \cite{Wilkins_2016} and \cite{Yung_2020} showed that accounting for binary stellar populations results in production efficiencies that are higher by $\sim0.2$ dex, and hence a better match to observed values. Binary evolution pathways are an important source of ionising photons \citep{Ma_2016, Eldridge_2020}. Processes such as stripping, mass transfer, and mergers result in prolonged Lyman-continuum (LyC) emission compared to single star populations \citep{Eldridge_2008, Stanway_2016, Gotberg_2019}. In addition, massive stars with low metallicities undergo quasi-homogeneous evolution, in which the rotational mixing that results from mass transfer leads to higher surface temperatures -- and hence stronger LyC emission \citep{Stanway_2016}.

In this work, we use the First Light and Reionisation Epoch Simulations \citep[\flares][]{Lovell_2021, Vijayan_2021} to make predictions for the ionising emissivity, specific ionising emissivity, and ionising photon production efficiency of galaxies with $M_{\star}\gtrsim10^8\ {\rm M_{\odot}}$ at $z=5-10$. \flares\ is a suite of high-redshift hydrodynamic zoom simulations, run using the \eagle\ subgrid physics model \citep{Crain:2015, Schaye:2015}. In the zoom simulation method, regions are selected from a dark matter-only (DMO) parent box and resimulated with hydrodynamics. By sampling a wide range of overdensities, and doing so with emphasis on highly overdense regions, we are able to efficiently simulate a large number of galaxies from a variety of environments. We note that in our analysis, the regions are weighted so as to recover the correct distribution of environments in the universe. This results in a galaxy sample with a wide range of properties that better represents the distribution of galaxies in our actual universe -- essential for studying trends in galaxy properties, and making predictions for observations, particularly those from JWST \citep[e.g.][]{Lovell_2022, Roper_2022, Wilkins_Colour_2022, Thomas_2023}.

The paper is structured as follows: in Section \ref{sec:theory}, we use a toy model to explore how the specific emissivity and production efficiency are affected by star formation, metal enrichment, SPS model and initial mass function (IMF). In Section \ref{sec:sims}, we introduce \flares\ and explain our methodology. Sections \ref{sec:spec_nion}, \ref{sec:ippe} and \ref{sec:nion} contain our analysis of the specific ionising emissivity, ionising photon production efficiency and ionising emissivity respectively. Finally, we summarise our findings in this work and present our conclusion in Section \ref{sec:conclusion}.

Throughout this work, we use the word `ionising' to mean `hydrogen-ionising'. Unless otherwise stated, we focus on stellar emission, and leave the study of AGN emission to a separate work that will contain a more in-depth analysis of AGN in \flares\ (Kuusisto et al., in prep). The terms `emissivity' and `production efficiency' are sometimes used interchangeably with `ionising emissivity' and `ionising photon production efficiency', respectively.

%% file: sections/2.theory.tex
\section{Theoretical Background}\label{sec:theory}

In this section, we use a simple model to explore theoretical predictions for the specific ionising emissivity, i.e. the rate at which ionising photons are produced by a stellar population per unit stellar mass ($\dot{N}_{\rm ion,intr}/M_\star$), and the ionising photon production efficiency \xiion. Figures \ref{fig:theory_sfh}--\ref{fig:theory_imf} show how these two properties are impacted by star formation history (SFH), metallicity, choice of SPS model, and choice of IMF. Note that Figures \ref{fig:theory_sfh} and \ref{fig:theory_Z} were made using v2.2.1 of the Binary Population And Stellar Synthesis (BPASS) SPS library \citep[][the default choice in \flares]{Stanway_2018}.

\subsection{Star formation and metal enrichment history}\label{sec:theory_sfzh}

\begin{figure}
	\includegraphics[width=\columnwidth]{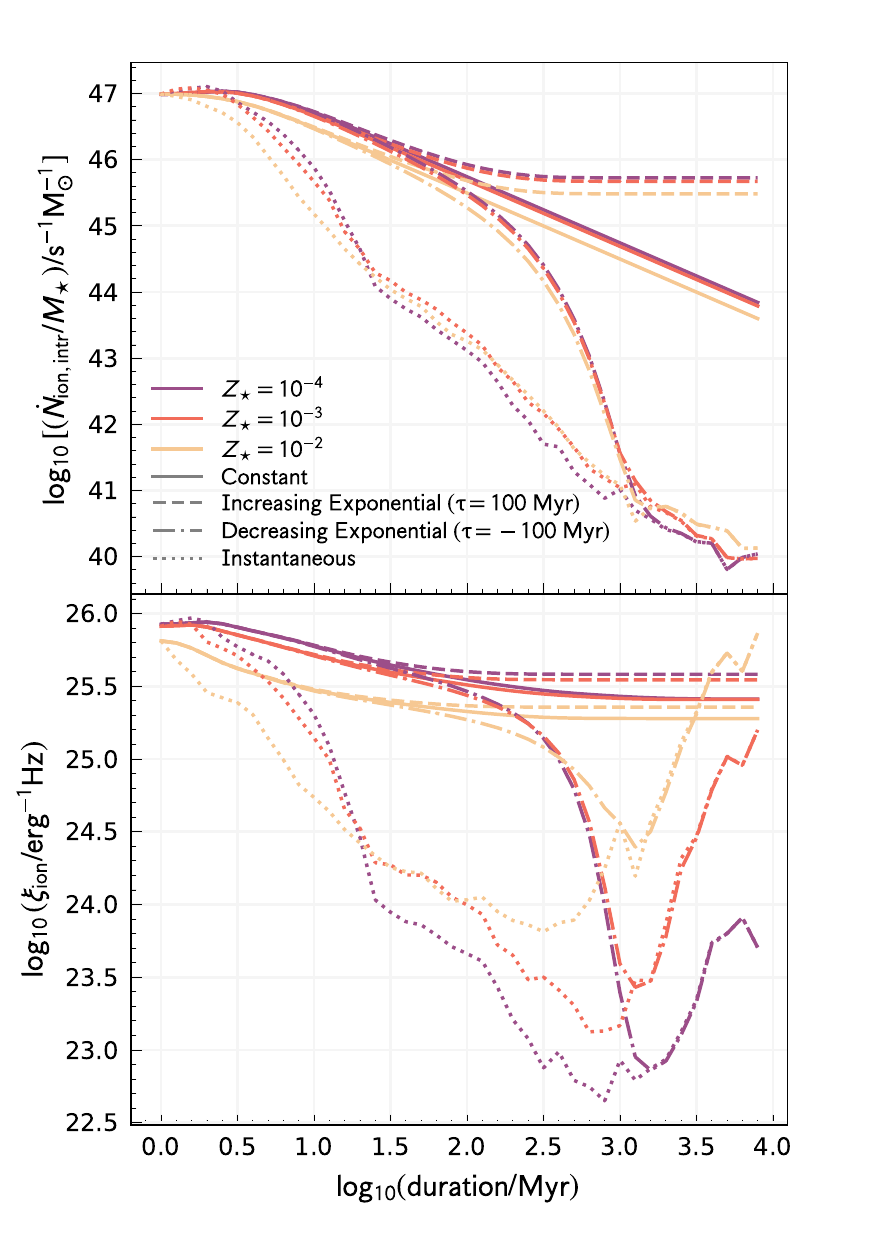}
	\caption{The specific ionising emissivity \nionM\ (top) and production efficiency (bottom) as a function of star formation duration for a range of star formation history parameterisations and metallicities. The solid, dashed, dot-dashed, and dotted lines denote constant, increasing exponential, decreasing exponential, and an instantaneous burst of star formation, respectively. \label{fig:theory_sfh}}
\end{figure}

Hot, massive stars are the main source of ionising photons in a galaxy. As these massive stars have very short lifespans, the total ionising emissivity of a stellar population declines steeply as it ages. With this in mind, we can view the specific emissivity and production efficiency as reflections of the proportion of young stars in a stellar population. Figure \ref{fig:theory_sfh} shows how different star formation histories affect the specific emissivity and production efficiency as a function of star formation duration. Galaxies with a constant star formation history have a continuously increasing population of old stars, while young, massive stars are formed at the same rate. Hence, the proportion of young stars decreases over time in a constant star formation model, and we observe a corresponding decrease in the specific emissivity and production efficiency. In this model, the production efficiency plateaus after a certain point in time, due to the decrease in far-UV emission as stars age. For galaxies with an exponentially increasing SFH, a plateau is observed for the specific emissivity as well, as the rapidly increasing population of young stars produces enough ionising radiation to balance the increase in stellar mass. As for galaxies with an exponentially decreasing SFH, the sharp drop in the number of young stars leads to a steep drop in both the production efficiency and the specific emissivity at $\sim10^{2.5}$ Myr. For both the decreasing exponential and instantaneous SFH, we observe an increase in the production efficiency after $\sim1$ Gyr. This is likely due to the growing population of white dwarfs. Hot white dwarfs provide an additional source of LyC emission at late times. Due to their lower luminosity, they make a smaller contribution to the total far-UV emission than the most massive surviving Main Sequence stars, hence the increase in production efficiency. The upturn is not observed in the case of a constant or increasing SFH because the contribution of white dwarfs to the total LyC emission is relatively small. Figure \ref{fig:theory_sfh} also shows the effect of metallicity for a given star formation history. On the whole, lower metallicities lead to higher values of the specific emissivity and production efficiency.

\begin{figure}
	\includegraphics[width=\columnwidth]{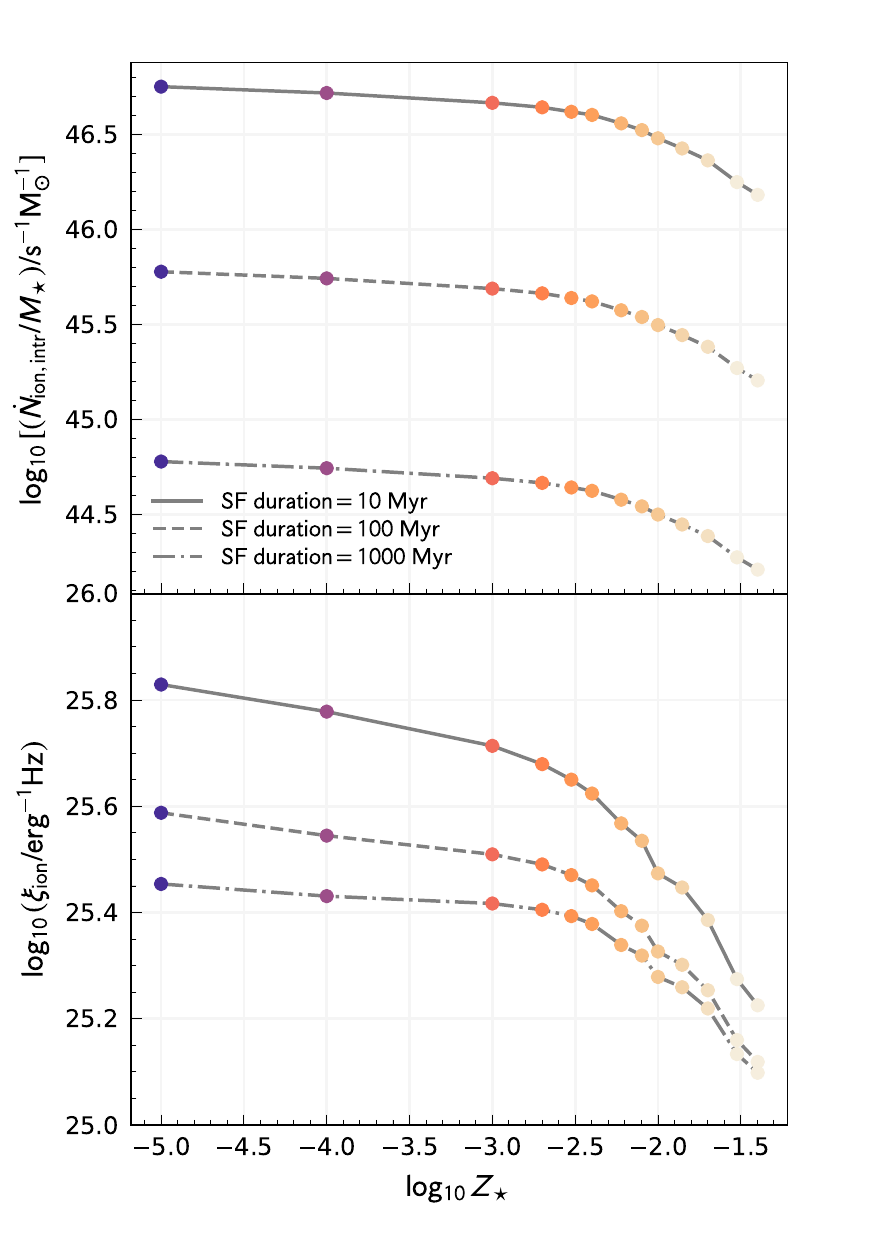}
	\caption{The specific ionising emissivity (top) and production efficiency (bottom) as a function of stellar metallicity for different star formation histories. The solid, dashed, and dot-dashed lines represent galaxies that have experienced 10, 100, and 1000 Myr of constant star formation respectively. \label{fig:theory_Z}}
\end{figure}

This is shown more clearly in Figure \ref{fig:theory_Z}, where we plot the two properties as a function of metallicity for three different durations of continuous star formation. This reveals a strong dependence on metallicity at $Z_\star>0.001$, with a weaker trend at lower metallicities. From $Z_\star=0.001\to 0.01$, the specific emissivity and production efficiency drop by $\approx 0.2$ dex. Metals enable more efficient cooling and by the same physical process increase the opacity of stars, hence an increase in metallicity leads to stars having lower surface temperatures and consequently a lower production rate of ionising photons. The impact of different star formation durations can be attributed to varying compositions of stellar ages, as mentioned in the discussion above. A galaxy that has only been forming stars for 10 Myr would have a higher proportion of young, massive stars than a galaxy that has been forming stars for the last 100 Myr, leading to a higher specific emissivity and production efficiency.

\subsection{Choice of stellar population synthesis model}

\begin{figure}
	\includegraphics[width=\columnwidth]{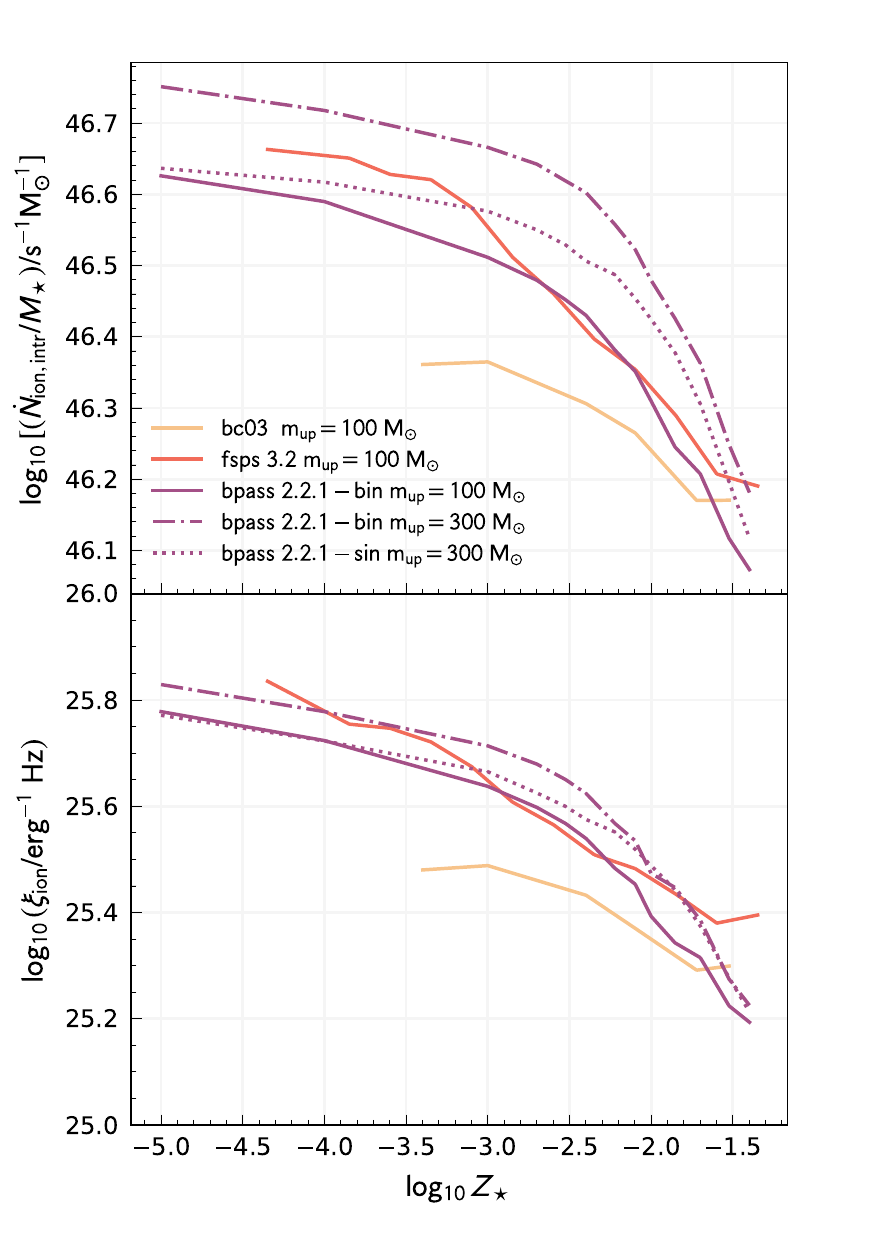}
	\caption{The specific ionising emissivity (top) and production efficiency (bottom) as a function of stellar metallicity, assuming 10 Myr of constant star formation. This is shown for different SPS models (BPASS v2.2.1, BC03, FSPS v3.2). The solid lines represent models with an upper-mass limit of 100 ${\rm M_{\odot}}$. The binary BPASS model with an upper-mass limit of 300 ${\rm M_{\odot}}$ is shown by the dot-dashed line, while the single-star BPASS model with an upper-mass limit of 300 ${\rm M_{\odot}}$ is shown by the dotted line. \label{fig:theory_sps}}
\end{figure}

While we do not explore changing the SPS model in \flares, it is useful to consider the impact that this would have using our simple toy model. Figure \ref{fig:theory_sps} shows the specific emissivity and production efficiency as a function of metallicity for three different SPS models: Binary Population And Stellar Synthesis (BPASS) v2.2.1 \citep{Stanway_2018}, Flexible Stellar Population Synthesis (FSPS) v3.2 \citep{FSPS_2009, FSPS_2010}, and BC03 \citep{BC03}. In each case, we assume 10 Myr constant star formation. Considering an upper-mass limit of $100\,\rm{M_{\odot}}$, we find that BPASS and FSPS yield comparable specific emissivities and production efficiencies at typical \flares\ metallicities (0.001-0.01). Outside of this metallicity range, using FSPS results in higher values of the production efficiency than BPASS, while BC03 consistently assigns lower values of the production efficiency except at the highest metallicities.

The effect of binaries can be seen by comparing the single-star and binary BPASS models with an upper-mass limit $m_{\rm up}=300\,\rm{M_{\odot}}$. (We note here that FLARES uses a similar binary BPASS model with $m_{\rm up}=300\,\rm{M_{\odot}}$). The inclusion of binary systems has a greater impact at lower metallicities, boosting the specific emissivity by $\sim0.1$ dex and the production efficiency by $\sim0.05$ dex at $Z_\star<10^{-2.5}$. In a similar analysis, \cite{Shivaei_2018} compared binary and single-star BPASS (v2) models and found that for galaxies with 300 Myr constant star formation and a metallicity of $Z_\star=10^{-2.7}$, the presence of binaries increases the production efficiency by 0.17 dex.


\begin{figure}
	\includegraphics[width=\columnwidth]{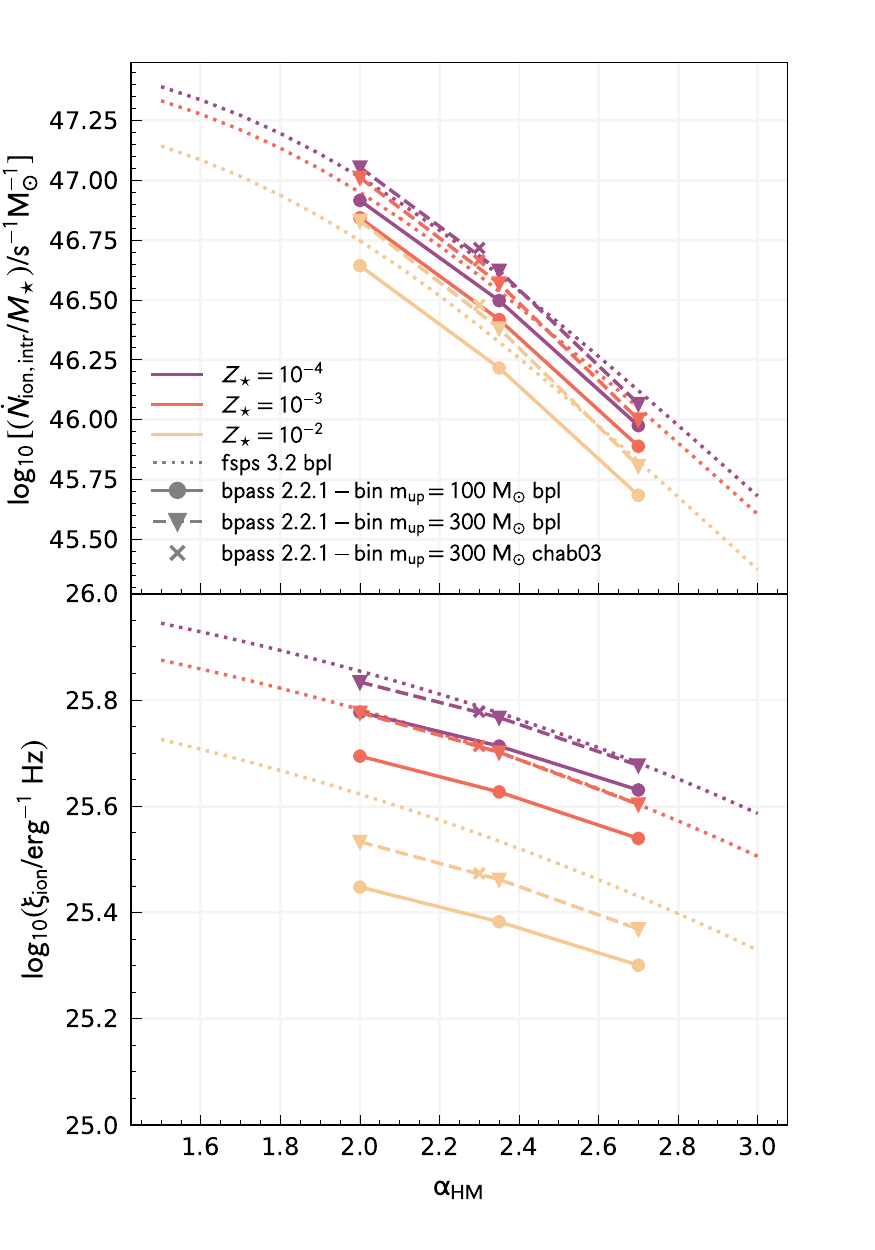}
	\caption{The specific ionising emissivity (top) and production efficiency (bottom) as a function of the high-mass slope for a range of metallicities and SPS models, using either a broken power law (BPL) or Chabrier \protect\citep[Chab03,][]{Chabrier_2003} IMF. \label{fig:theory_imf}}
\end{figure}

\subsection{Initial mass function}

In Figure \ref{fig:theory_imf}, we show the specific emissivity and production efficiency as a function of the high-mass slope $\alpha_{\rm HM}$ for both FSPS and BPASS. In both models, $\alpha_{\rm HM}$ is the slope at $>1\, {\rm M_{\odot}}$. However, the models have a different behaviour at low-masses, explaining some of the offset between the two. Most of the examples in Figure \ref{fig:theory_imf} adopt a broken power law (BPL) IMF. We also include a binary BPASS model with a \cite{Chabrier_2003} IMF, as this is the form used in \flares. We find it gives a comparable result to the BPL BPASS model with the same upper-mass limit and $\alpha_{\rm HM}=2.35$ (see \cite{Stanway_2018} for more detail on the BPASS IMFs used).

Unsurprisingly, flattening the high-mass slope, and thus boosting the fraction of high-mass stars, yields higher emissivities. Flattening the slope by $\Delta\alpha_{\rm HM}=0.5$ increases the specific emissivity by 0.5 dex. Since flattening the slope also boosts the UV luminosity, the impact on the production efficiency is weaker, $\sim0.2-0.3$ dex. For BPASS, we also consider two upper-mass limits: $m_{\rm up}=100$ and 300 ${\rm M_{\odot}}$. Extending the mass-range of stars to 300 ${\rm M_{\odot}}$ boosts the production efficiency by $\approx 0.15$ dex. A very similar result was found by \cite{Shivaei_2018}, who used v2 of the single-star BPASS model and found that increasing $m_{\rm up}$ from 100 to 300 ${\rm M_{\odot}}$ increased the production efficiency by 0.18 and 0.12 dex, assuming a high-mass slope of $\alpha_{\rm HM}=2.0$ and $\alpha_{\rm HM}=2.7$ respectively.

%% file: sections/3.sims.tex
\section{Methods}\label{sec:sims}

In this section, we introduce \flares, the suite of simulations used in this study. We define the physical properties used in this paper and describe the forward modelling procedure for obtaining galaxy SEDs from these physical properties.

\subsection{First Light And Reionisation Epoch Simulations}

The First Light And Reionisation Epoch Simulations (\flares) is a suite of hydrodynamical zoom simulations that probes galaxy formation and evolution at high redshift \citep{Lovell_2021, Vijayan_2021}. It consists of 40 spherical resimulations, each with a radius of 14$h^{-1}$Mpc. The regions for resimulation are selected from a (3.2 cGpc)$^{3}$ parent dark matter-only (DMO) simulation, the same as that used in the \ceagle\ zoom simulations \citep{Barnes_2017}. The large volume of the parent box provides access to a wide variety of environments. Taking advantage of this, the resimulated regions in \flares\ span an overdensity range of $\delta=-0.497\rightarrow0.970$ \cite[see Table A1 of][]{Lovell_2021}, with a bias towards highly overdense environments, where massive galaxies are more likely to form \citep{Chiang_2013, Lovell_2018}. The regions are selected at $z=4.67$, when the most extreme overdensities are only mildly non-linear, so as to preserve the rank ordering of the overdensities at high redshift. When studying galaxy population statistics, it is necessary to recreate the original distribution of environments in the parent simulation. To do so, we use a weighting scheme that reduces the contribution from rarer regions, i.e. the most underdense and overdense ones (see \cite{Lovell_2021} for a more detailed explanation). Throughout this work, aggregate values such as the median are obtained by applying this weighting scheme.

The \flares\ regions are resimulated with hydrodynamics using the AGNdT9 variant of the \eagle\ subgrid physics model \citep{Schaye:2015, Crain:2015}. The AGNdT9 variant was chosen as it produces similar mass functions as the fiducial model, while providing a better match to observations of the hot gas properties in groups and clusters \citep{Barnes_2017_CEagle}. \flares\ has an identical resolution to the fiducial \eagle\ model: dark matter and initial gas particles are of mass m$_{\mathrm{dm}}=9.7\times10^6$ M$_{\odot}$ and m$_{\mathrm{g}}=1.8\times10^6$ M$_{\odot}$ respectively, with a softening length of $2.66\, \mathrm{ckpc}$. The output of the simulations is stored at integer redshifts at $z=5-15$.

\subsection{Measuring galaxy properties}

As with the \eagle\ simulation, galaxies in \flares\ are first grouped using the Friends-Of-Friends (FOF) algorithm \citep{Davis_1985} before being identified as substructures using the \subfind\ algorithm \citep{Springel_2001}. All galaxy properties in this paper are measured using particles that lie within a 30 pkpc (physical kpc) aperture of the most bound particle in each substructure. 

We define the age of a galaxy as the initial mass-weighted median age of its constituent stellar particles (i.e. within the 30 pkpc aperture). Metallicity is similarly defined as the initial mass-weighted median metallicity of a galaxy's stellar particles. The star formation rate (SFR) is calculated by considering the total stellar mass formed over the most recent 50 Myr, and the specific star formation rate (sSFR) is the star formation rate per unit stellar mass.

When studying the physical properties of galaxies, we limit our analysis to galaxies with a stellar mass of $M_*>10^8\,\mathrm{M_\odot}$, as this is the mass range resolved by \flares. When considering the observational properties of galaxies, we add an additional cut in the dust-attenuated far-UV luminosity: $L_{\rm FUV,\,dust}>10^{28}\,\mathrm{erg\,s^{-1}\,Hz^{-1}}$ (this corresponds to a cut at $M_{\rm FUV}=-18.4$).

\subsection{SED Modelling}\label{sec:sed}

Here we provide a brief description of the stellar SED modelling in \flares, and refer the reader to \cite{Vijayan_2021} for a more in-depth explanation. The SED modelling procedure for AGN is detailed in Kuusisto et al. (in prep).

\subsubsection{Stellar SED}

To begin, we assign a stellar SED to each stellar particle, according to its mass, age and metallicity. We use v2.2.1 of the Binary Population And Spectral Synthesis \citep[BPASS,][]{Stanway_2018} SPS library, and assume a \cite{Chabrier_2003} IMF. As discussed in Section \ref{sec:theory}, the SPS model and IMF used have a strong influence on the ionising properties of galaxies.

The intrinsic emissivity $\dot{N}_{\rm ion,intr}$ and intrinsic $L_{\rm UV}$ used to derive \xiion\ in Equation \ref{eq:xiion} are both obtained from pure stellar SEDs.

\subsubsection{Nebular emission}\label{sec:neb}

Nebular emission occurs when LyC radiation from stars is reprocessed by gas and dust. We model birth clouds by associating each young stellar particle \citep[$<10$ Myr, assuming birth clouds disperse on these timescales;][]{Charlot_2000} with an ionisation-bounded $\rm H_{II}$ region. Nebular emission lines are obtained by using the SED of the stellar particle as the incident radiation field in version 17.03 of the \cloudy\ photoionisation code \citep{Ferland_2017}. We assume a solar abundance pattern, a covering fraction of 1 (equivalent to $f_{\rm esc}=0$), a birth cloud metallicity identical to that of the stellar particle, and a hydrogen density of ${\rm log}_{10}(n_{\rm H}/{\rm cm}^{-3})=2.5$. Dust depletion factors and relative abundances are taken from \cite{Gutkin_2016}. We use a metallicity- and age- dependent ionisation parameter $U$, scaled from a reference value of $U_{\rm ref}=10^{-2}$ at $Z_{\rm \star,ref}=0.02$ and $t_{\rm ref}=1$ Myr. For a spherical ionised region around a stellar particle of metallicity $Z_\star$ and age $t$, $U$ scales with the intrinsic ionising emissivity $\dot{N}_{\rm ion}$ of the stellar particle as follows:
\begin{equation}
    U(Z_\star,t)=U_{\rm ref}\left(\frac{\dot{N}_{\rm ion}(Z_\star,t)}{\dot{N}_{\rm ion,ref}(Z_{\rm \star,ref},t_{\rm ref})}\right)^{1/3}.
\end{equation}
$\dot{N}_{\rm ion, ref}$ is the reference emissivity, obtained for a stellar particle with metallicity $Z_{\rm \star,ref}$ and age $t_{\rm ref}$. We refer the interested reader to Section 2.1.2 of \cite{Wilkins_2023_OIII} for a more comprehensive explanation.

\subsubsection{Dust attenuation}

There are two components to our dust model: the first accounts for dust extinction in the birth cloud, and the second accounts for dust extinction in the intervening ISM.

As with modelling nebular emission, we associate young stellar particles ($<10$ Myr) with a birth cloud. The birth cloud dust optical depth in the V-band, $\tau_{\rm BC, V}$, is taken to be metallicity-dependent:
\begin{equation}
    \tau_{\rm BC, V}=\kappa_{\rm BC}(Z_\star/0.01),
    \label{eq:tau_bc}
\end{equation}
where $Z_\star$ is the metallicity of the stellar particle, and $\kappa_{\rm BC}$ is a normalisation factor with a value of $1$. For older stellar particles ($>10$ Myr), $\tau_{\rm BC, V}=0$.

To model dust extinction in the intervening ISM, we treat each stellar particle as a point-like particle and take our line of sight (LOS) to be the $z$-axis. We first obtain the LOS metal column density $\Sigma(x,y)$ by integrating the density field of gas particles along the LOS, using the smoothed particle hydrodynamics (SPH) smoothing kernel of the gas particles. The column density is then converted to the ISM dust optical depth in the V-band:
\begin{equation}
    \tau_{\rm ISM, V}(x, y)={\rm DTM}\;\kappa_{\rm ISM}\;\Sigma(x,y),
\end{equation}
where DTM is the dust-to-metal ratio, obtained for each galaxy using a fitting function dependent on stellar age and gas-phase metallicity \citep[Equation 15 in][]{Vijayan_2019}. $\kappa_{\rm ISM}$ is a normalisation factor with a value of 0.0795, chosen to match the $z=5$ UV luminosity function in \cite{Bouwens_2015_UVLF}.

The optical depth at other wavelengths is given by an inverse power law:
\begin{equation}
    \tau_\lambda=(\tau_{\rm BC, V}+\tau_{\rm ISM, V}(x, y))\times(\lambda/550{\rm nm})^{-1}.
\end{equation}
This expression can then be applied to the stellar particle SEDs. Comparing our model with the Small Magellanic Cloud (SMC) \citep{Pei_1992} and Calzetti \citep{Calzetti_2000} curves, we find that the dust model detailed here results in very similar values of attenuation in the far-UV. However, measurements of the UV-continuum slope $\beta$ are more sensitive to the choice of attenuation curve. The \flares\ dust model is flatter in the UV than the SMC curve, though not as much as the Calzetti curve, leading to values of $\beta$ in between the two curves, but closer to the SMC values \citep[see Section C1 of the Appendix in][]{Vijayan_2021}.

%% file: sections/4.ippe_mstar.tex
\begin{figure}
	\includegraphics[width=\columnwidth]{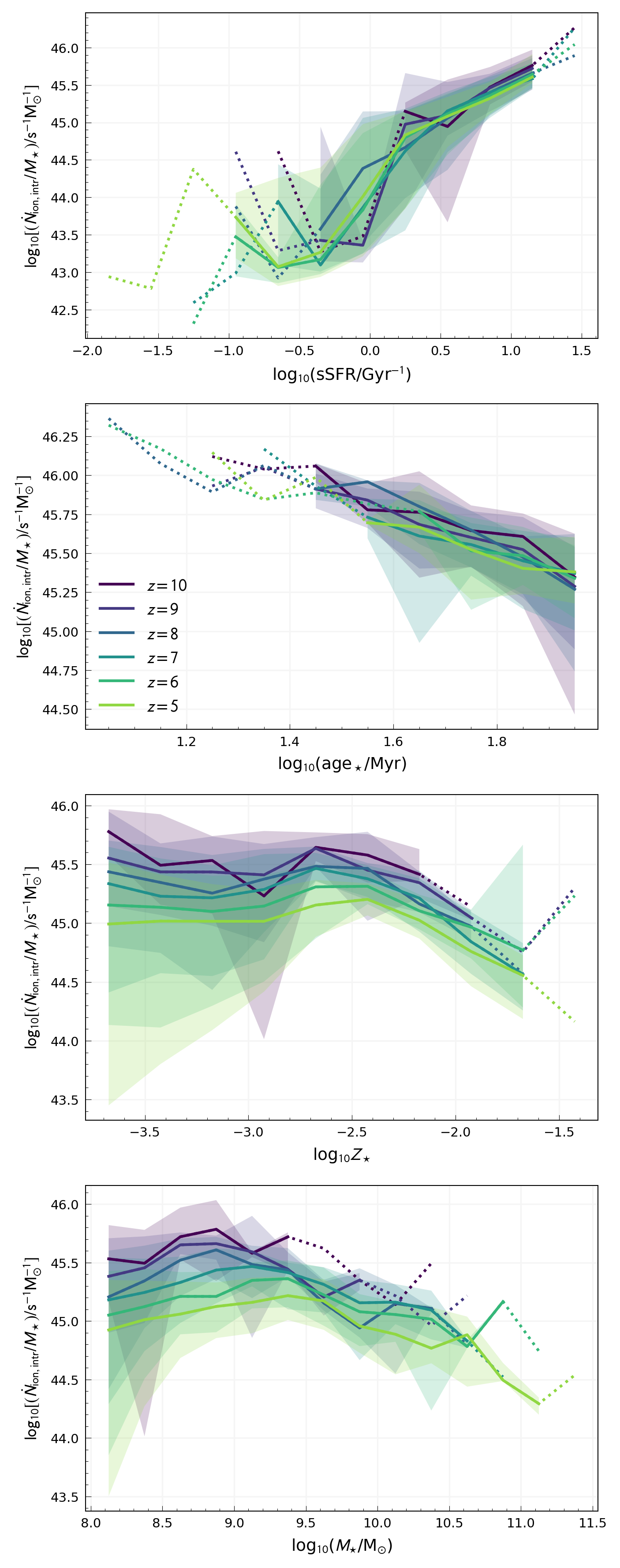}
	\caption{Specific ionising emissivity as a function of (from top to bottom): specific star formation rate, stellar age, stellar metallicity, and stellar mass. Trend lines show the weighted median specific ionising emissivity, and are coloured by redshift. Dotted lines are used to represent bins containing fewer than 10 galaxies. Shaded regions denote the 16-84th percentile range. \label{fig:ippe_mstar_col}}
\end{figure}

\begin{figure}
	\includegraphics[width=\columnwidth]{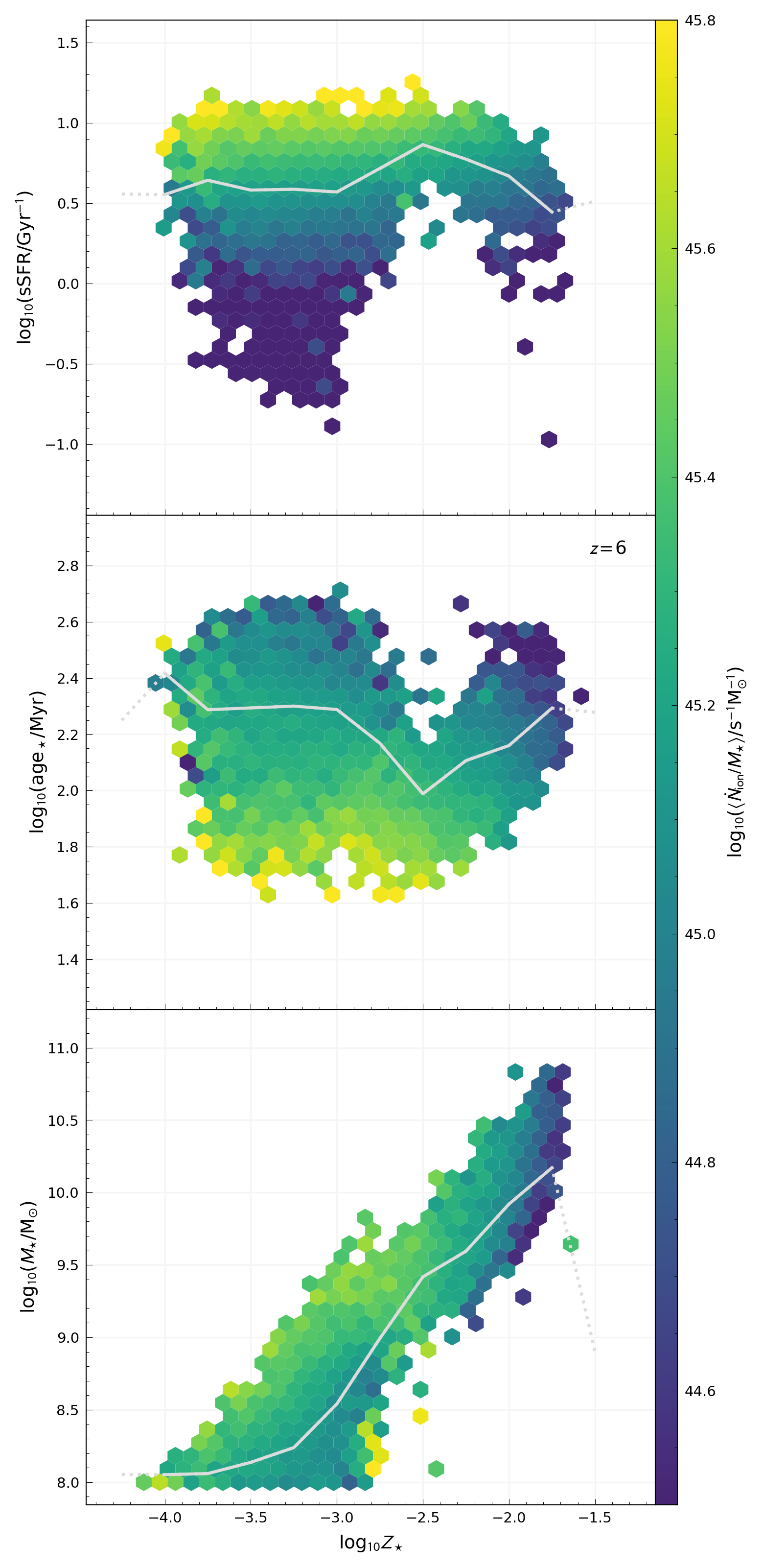}
	\caption{(From top to bottom) Stellar age, specific star formation rate and stellar mass as a function of stellar metallicity at $z=6$. Similar trends are observed at other redshifts. The hex bins are coloured by mean specific ionising emissivity (only bins containing two or more galaxies are displayed). The weighted median line in grey accounts for all galaxies, including those not displayed in a hex bin. \label{fig:ippe_mstar_hex_col}}
\end{figure}

\section{Specific ionising emissivity}\label{sec:spec_nion}

In this section, we study the dependence of the specific ionising emissivity, defined as the intrinsic ionising emissivity per unit stellar mass ($\dot{N}_{\rm ion,intr}/M_\star$), on the following physical properties: specific star formation rate, age, metallicity, and stellar mass. \cite{Wilkins_2023_SFZH} have studied the star formation and metal enrichment histories of galaxies in \flares\ in depth. Here, we link these quantities to the specific emissivity. We focus our analysis on galaxies with $M_{\star}>10^8\ {\rm M_{\odot}}$.

\subsection{Star formation and metal enrichment history}\label{sec:sfzh}

The first panel of Figure \ref{fig:ippe_mstar_col} shows a strong positive correlation between specific emissivity and specific star formation rate. As mentioned in Section \ref{sec:theory_sfzh}, it is the young, massive, short-lived stars that produce large amounts of ionising radiation. A high specific star formation rate is indicative of a large fraction of these young stars. As such, the ionising emissivity of a galaxy essentially traces recent star formation. A tighter relation would be observed if we were to use a star formation rate defined on a shorter timescale, due to SFH variability. 

The negative correlation of specific emissivity with age (second panel of Figure \ref{fig:ippe_mstar_col}) can be explained along similar lines. Since we define age as the initial mass-weighted median age of the stellar population, the age of a galaxy tells us if its stellar population is generally young or old. The younger the stellar population, the more efficient a galaxy is at producing ionising radiation per unit mass. The effect of age is also reflected in the redshift dependence of the specific emissivity, most obvious in the third and fourth panels of Figure \ref{fig:ippe_mstar_col}, where we see smaller values of the specific emissivity at lower redshifts, for a given metallicity or stellar mass. In the case of stellar mass, the specific emissivity decreases by $\sim0.1$ dex between integer redshifts.

The relationship between specific emissivity and metallicity is shown in the third panel of Figure \ref{fig:ippe_mstar_col}. At extremely low metallicities, up until $Z_\star\sim10^{-3}$, the trend is more or less flat, with at most a weak relationship of decreasing specific emissivity with increasing metallicity. Past $Z_\star\sim10^{-3}$, the specific emissivity increases and peaks at $Z_\star\sim10^{-2.5}$, before decreasing steeply. In the top panel of Figure \ref{fig:theory_Z}, which shows how the specific emissivity evolves with metallicity for a simple model, we see a similar distinction between trends at high and low metallicity. However, the peak is not observed in Figure \ref{fig:theory_Z}, as it results from the more complex stellar populations of galaxies in \flares. \cite{Roper_2023} investigated the size evolution of galaxies in \flares\ and found that galaxies around a stellar mass of $M_\star\sim10^{9.5}\,\rm{M_\odot}$ tend to undergo a burst of star formation in their cores, triggered by enriched gas cooling to higher densities. This leads to an increased specific star formation rate -- which in turn causes the specific emissivity to increase. Since galaxies in \flares\ exhibit a strong mass-metallicity relation (see bottom panel of Figure \ref{fig:ippe_mstar_hex_col} for an example at $z=6$), we find that the burst of star formation occurs around a particular metallicity range as well. The turnover in the specific emissivity likely occurs for a few different reasons. The top panel of Figure \ref{fig:theory_Z} shows that the specific emissivity of a stellar population decreases more steeply at $Z_\star>10^{-2.5}$, as higher metallicities lead to cooler stars. There is also the role of feedback in regulating star formation -- these metal-rich galaxies tend to be more massive, and likely have their star formation regulated by AGN feedback. Figure \ref{fig:ippe_mstar_hex_col} presents the distribution of the physical properties of \flares\ galaxies at $z=6$, coloured by specific emissivity. We note that the dip in age and the peak in specific star formation rate observed at $Z_\star\sim10^{-2.5}$ are associated with the aforementioned burst of star formation.

The top and middle panels of Figure \ref{fig:ippe_mstar_hex_col} show how the specific emissivity is largely dictated by specific star formation rate and age. The effect of metallicity is more subtle, only becoming evident at higher metallicities ($Z_\star>10^{-2.5}$), where we see a slight decrease in the specific emissivity for constant values of age or specific star formation rate.

\subsection{Stellar mass}

In the lowermost panel of Figure \ref{fig:ippe_mstar_col}, we observe a general trend of decreasing specific emissivity at high stellar masses ($M_\star>10^9\;\rm{M_\odot}$). This is primarily due to the effect of metallicity -- galaxies in \flares\ exhibit a strong mass-metallicity relation, as shown in the bottom panel of Figure \ref{fig:ippe_mstar_hex_col} \cite[for other redshifts, see Figure 2 of][]{Vijayan_2021}. The median age of galaxies is also seen to increase slightly at high stellar masses \cite[Figure 5 of][]{Wilkins_2023_SFZH}, which would also contribute to the trend. At lower stellar masses, the trend is flatter, with a tentative negative slope.

%% file: sections/5.ippe_fuv.tex
\begin{figure*}
	\includegraphics[width=2\columnwidth]{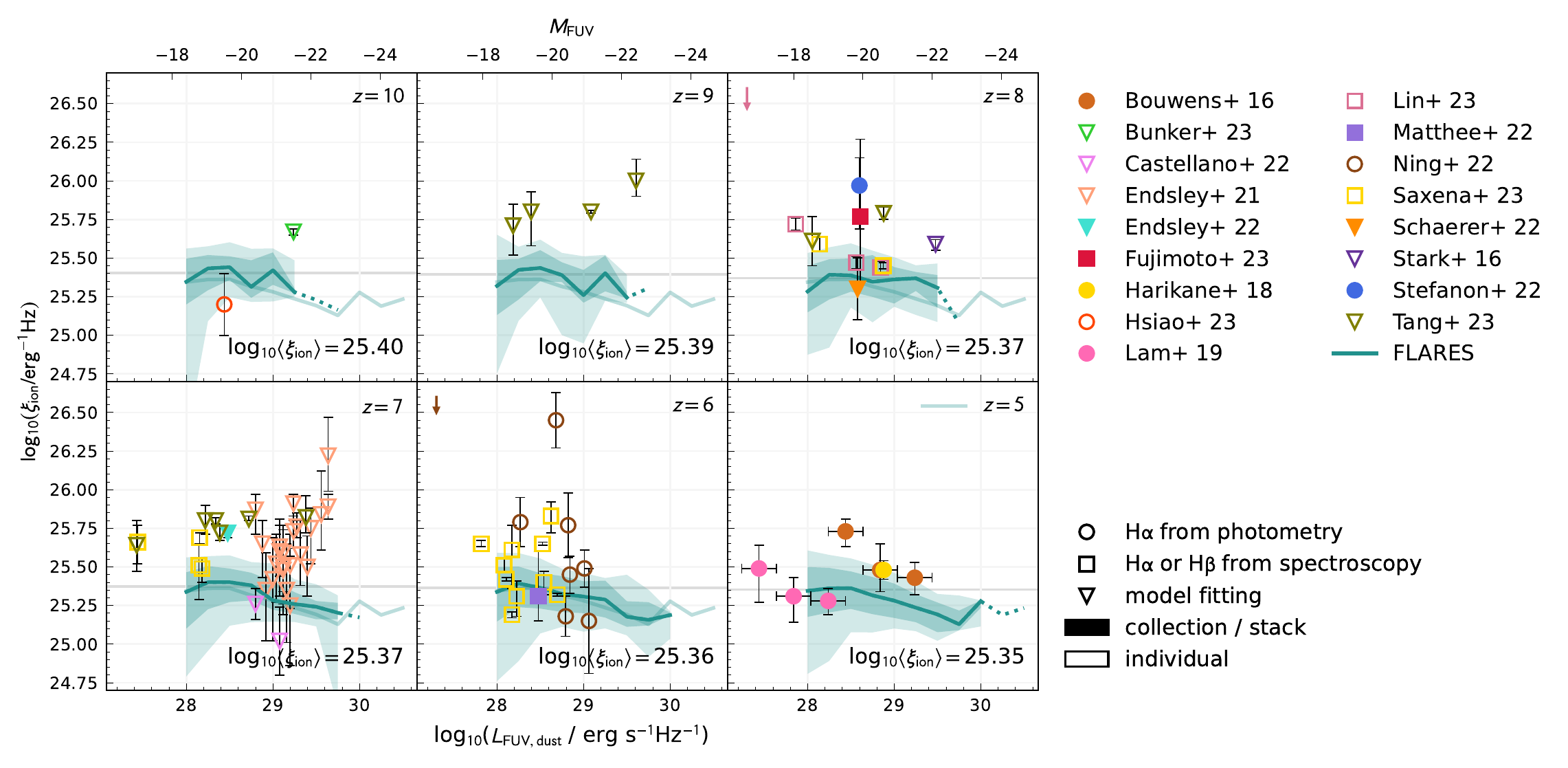}
	\caption{Ionising photon production efficiency as a function of the dust-attenuated far-UV luminosity (rest frame 1500\AA), for redshifts $z=5-10$. Trend lines show the weighted median ionising photon production efficiency in \flares, and shaded regions denote the 1 and 2$\sigma$ range. The faint, horizontal grey line indicates the weighted mean of the sample at each redshift. The translucent blue line plotted across all panels shows the weighted median at $z=5$. Observations are displayed as scatter points: those with a transparent fill are measurements of individual galaxies; those with a solid fill are aggregated values, representing either stacks or collections of galaxies; circular and square data points represent measurements of the production efficiency obtained using Balmer emission line fluxes from photometry and spectroscopy respectively; triangular data points represent measurements of the production efficiency obtained from model fitting (this is a broad term that encompasses SED fitting). Observations are plotted in the panel corresponding to the nearest integer redshift. Arrows indicate how values may change when accounting for dust \protect\citep[for measurements by][]{Ning_2022, Lin_2023}. A version of this plot with \xiion\ as a function of stellar mass can be found in the Appendix (Figure \ref{fig:ippe_psfuv_mstar}). \label{fig:ippe_psfuv}}
\end{figure*}

\section{Ionising photon production efficiency}\label{sec:ippe}

In this section, we explore how the ionising photon production efficiency varies with observable properties, namely the far-UV luminosity, UV continuum slope and \oiii\ equivalent width. Since we are now making predictions in the observer space, on top of our mass cut we impose a luminosity cut using the dust-attenuated far-UV luminosity: $\log_{10}(L_{\rm FUV, dust}/\mathrm{erg\,s^{-1}Hz^{-1}})>28$ (this corresponds to a cut at $M_{\rm FUV}=-18.4$).

For most of this section, we define the production efficiency following Equation \ref{eq:xiion}, using the intrinsic far-UV luminosity to normalise the ionising emissivity. In Section \ref{sec:dm1}, we will discuss an alternate definition of the production efficiency, \xidust, that uses the dust-attenuated far-UV luminosity in place of the intrinsic value.

We note that estimating the production efficiency from line fluxes requires an assumption of the ionising photon escape fraction \fesc, as only reprocessed photons are responsible for nebular emission. For example, in the case of \ha, the intrinsic line luminosity can be expressed as \citep{Leitherer_1995}: 
\begin{equation}
    L_{\rm{H}\alpha}\rm{[erg\,s^{-1}]}=1.36\times10^{-12}(1-f_{\rm{esc}})\dot{N}_{\rm ion, intr}\rm{[s^{-1}]}.
\end{equation}
\hb\ and \hg\ fluxes can be used to obtain the intrinsic ionising emissivity $\dot{N}_{\rm ion, intr}$ in the same manner, applying the appropriate conversion factors. With \fesc\ being a highly uncertain parameter, we choose, where possible, to compare our results with measurements that assume $f_{\rm esc}=0$, essentially considering the lower bound of the production efficiency.

\subsection{High-z observations}

\begin{figure*}
	\includegraphics[width=2\columnwidth]{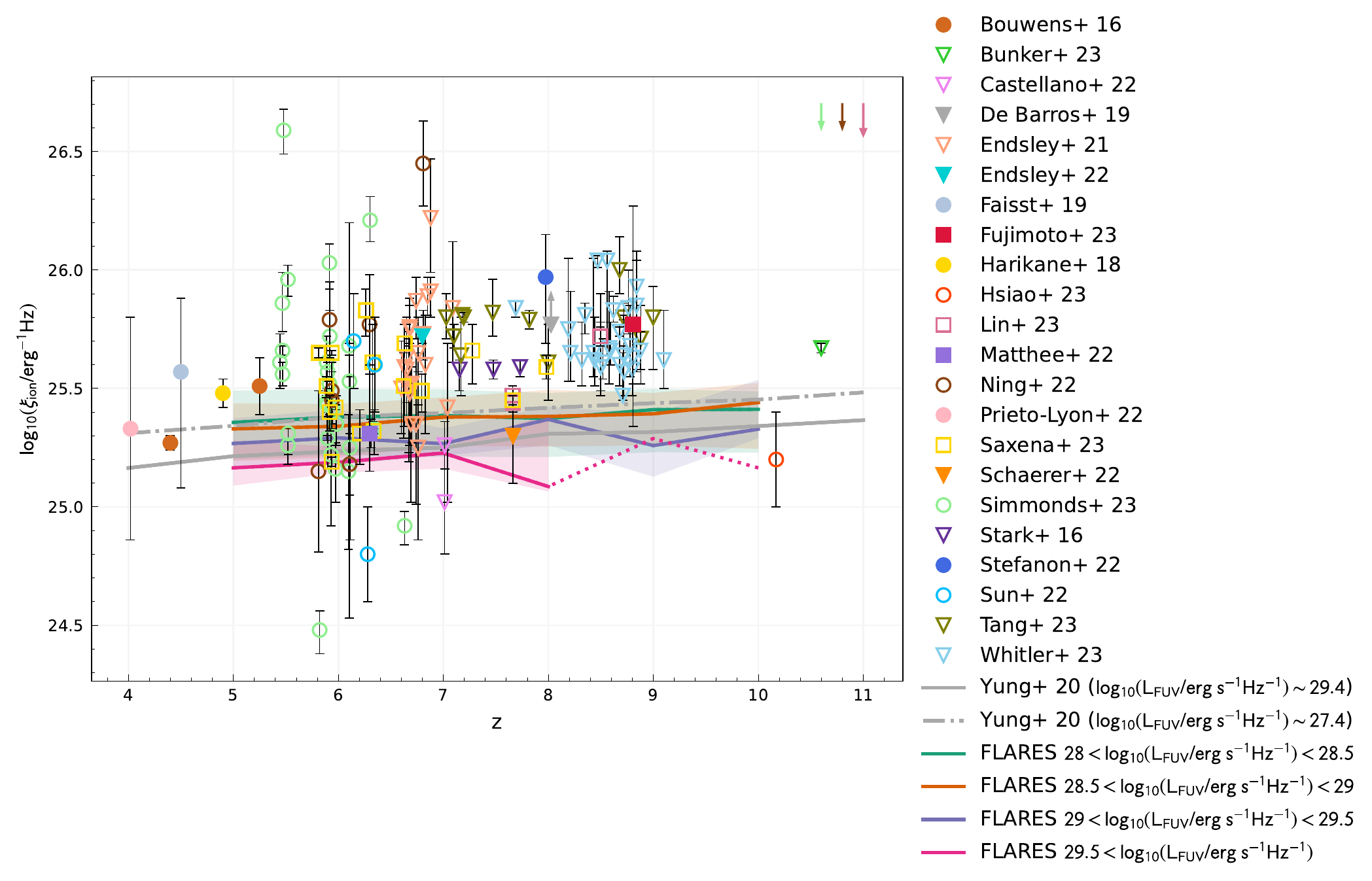}
	\caption{Ionising photon production efficiency as a function of redshift. Trend lines and their colour-associated shaded regions represent results from theoretical models: in grey are the median trends from \protect\cite{Yung_2020}, with the solid (dash-dotted) line corresponding to galaxies with of far-UV luminosity $\log_{10}(L_{\rm FUV}/\mathrm{erg\;s^{-1}Hz^{-1}})=29.4$ ($\sim27.4$); in the remaining colours are the weighted trends from \flares\ in bins of far-UV luminosity, with dotted lines representing bins with fewer than 10 galaxies, and the shaded regions denoting the $1\sigma$ range. Observations are shown as scatter points, with the same classification by marker style and fill as in Figure \ref{fig:ippe_psfuv}. Arrows in the top right corner indicate how values may change when accounting for dust \protect\citep[for measurements by][]{Ning_2022, Lin_2023, Simmonds_2023}. \label{fig:comparison_luv}}
\end{figure*}

Below, we list the high-redshift ($z>5$) observations of the production efficiency that we compare our results with (in Figures \ref{fig:ippe_psfuv}, \ref{fig:comparison_luv}, \ref{fig:ippe_beta}, \ref{fig:ippe_dm1fuv_compare} and \ref{fig:ippe_psfuv_mstar}). We have restricted our analysis to redshifts $z\gtrsim5$, as this is the range targeted by \flares:

\begin{itemize}
\item \cite{Bouwens_2016} measured \xiion\ for a sample of 22 galaxies at $z=5.1-5.4$, using \ha\ fluxes estimated from SPITZER/IRAC photometry. For the values plotted, the apparent \ha\ fluxes and UV-continuum were corrected for dust using measurements of the UV-continuum slope, assuming an SMC dust law \citep{Pei_1992}. In Figure \ref{fig:ippe_psfuv} (\ref{fig:ippe_psfuv_mstar}), the galaxies are binned by luminosity (stellar mass), with bin widths denoted by error bars.

\item \cite{Bunker_2023} analysed JWST/NIRSpec spectroscopy of GN-z11, a Lyman break galaxy with a derived redshift of $z=10.6$. Use of the Balmer lines and SED fitting with \beagle\ both lead to an estimate of $\log_{10}(\xi_{\rm ion}\rm{/erg^{-1}Hz})\approx25.7$. The calculation involving the Balmer lines does not include a dust correction, since the data suggests very little dust attenuation.

\item \cite{Castellano_2022} analysed VLT/X-SHOOTER observations of two $z\approx7$ \lya\ emitters (LAEs), thought to reside in a reionised bubble in the Bremer Deep Field (BDF). \xiion\ was obtained from a \beagle\ \citep{Chevallard_2016} fit to the spectroscopy and available photometry in the field, assuming an exponentially delayed SFH with a 10 Myr burst of constant star formation, and dust treatment following \cite{Charlot_2000} and \cite{Chevallard_2013}.

\item \cite{DeBarros_2019} analysed the Spitzer and HST photometry of $z\sim8$ galaxies and obtained median values of $\log_{10}(\xi_{\rm ion,dust}\rm{/erg^{-1}Hz})=26.07$ and 26.29 using SMC and Calzetti dust attenuation curves, respectively. Due to uncertain parameters in the SED fitting, the authors emphasise a lower limit of $\log_{10}(\xi_{\rm ion,dust}\rm{/erg^{-1}Hz})>25.77$ (indicated by an upwards arrow in Figures \ref{fig:comparison_luv} and \ref{fig:ippe_psfuv_mstar}), and note that the intrinsic \xiion\ should be very similar to \xidust, due to small dust attenuation.

\item \cite{Endsley_2021} studied 22 \oiiihb\ emitters at $z\sim7$ and obtained \xiion\ by fitting to SPITZER/IRAC photometry with the \beagle\ SED fitting tool, assuming an exponentially delayed SFH with an allowed recent ($<10$ Myr) burst, and an SMC dust law.

\item \cite{Endsley_2022} investigated 118 Lyman-break galaxies in the Extended Goth Strip (EGS) field. \xiion\ was obtained from a \beagle\ fit to JWST/NIRCam and HST/ACS photometry, assuming a constant star formation history (CSFH) and an SMC dust law. The authors find evidence for high specific star formation rates, in line with the high values of \xiion\ measured. 

\item \cite{Faisst_2019} measure \xiion\ for a collection of 221 galaxies at $z\sim4.5$, using the \ha\ line to obtain \nion. The \ha\ fluxes are estimated by comparing the measured Spitzer photometry to modelled colours from a variety of BC03 templates, focusing only on the optical continuum. We note that the calculation of \xiion\ assumes $f_{\rm esc}\sim0.1$. A wide range of values is obtained, between $\log_{10}(\xi_{\rm ion}\rm{/erg^{-1}Hz})=24.5-26.4$ (error bars in Figure \ref{fig:comparison_luv} show the 16th and 84th percentile values).

\item \cite{Fujimoto_2023} obtained \xiion\ for 7 $z\sim8-9$ galaxies with redshifts spectroscopically confirmed using JWST/NIRSpec spectroscopy. \xiion\ was measured using the \hb\ line, with dust extinction obtained from a fit to HST$+$JWST/NIRCam photometry and the \oiii$\lambda$5008 EW. The SED fitting was performed using \cigale, and assumed a delayed and final burst ($<$10 Myr) SFH, a Calzetti dust law for the stellar continuum and an SMC dust law for nebular emission.

\item \cite{Harikane_2018} measured \xiion\ for a stack of LAEs at $z=4.9$ (we have plotted the sample of 99 galaxies with $20<$\lya\ EW$<100$). To infer the \ha\ flux, the authors compared the stacked SED with a model SED, which was obtained from a \beagle\ fit assuming a CSFH and Calzetti dust curve. Taking into account the inferred escape fraction ($f_{\rm esc}=0.1$) of the sample would increase the production efficiency slightly, by $\sim0.05$ dex.

\item \cite{Hsiao_2023} studied JWST/NIRSpec spectroscopy of a $z\sim10.17$ galaxy (MACS0647-JD). \xiion\ was obtained using the \ha\ flux, which was estimated from the \hg\ emission line. The emission line flux and far-UV luminosity were not corrected for dust, however, results from SED fitting suggest little dust attenuation.

\item \cite{Lam_2019} measured \xiion\ for binned stacks of faint galaxies at $z\approx4-5$, using imaging data from HST and SPITZER/IRAC. The \ha\ flux used to calculate \xiion\ was obtained by fitting a model spectrum to the measured colours, and correcting for dust using an SMC dust law. The values plotted in Figure \ref{fig:ippe_psfuv} are binned by luminosity, with bin widths denoted by error bars.

\item \cite{Lin_2023} analysed the JWST/NIRSpec spectra of 3 lensed $z\sim8$ galaxies. We note that this work defined the production efficiency using the dust-attenuated \hb\ flux and the dust-attenuated far-UV luminosity. The arrows in Figures \ref{fig:ippe_psfuv} and \ref{fig:comparison_luv} roughly show how dust-attenuation, measured with the UV-continuum slope assuming an SMC dust law, would impact the production efficiencies ($\lesssim0.13$ dex lower). 

\item \cite{Matthee_2022} analysed a sample of $117$ \oiii\ emitters that were observed using JWST/NIRCam wide-field slitless spectroscopy. \xiion\ was obtained using the \hb\ line, with dust extinction inferred from the H$\gamma/$H$\beta$ ratio.

\item \cite{Ning_2022} used the \ha\ flux, estimated from JWST/NIRCam data, to measure \xiion\ for 7 spectroscopically confirmed Lyman break galaxies (LBGs) at $z\sim6$. We note that this work defined the production efficiency using the dust-attenuated \ha\ flux and the dust-attenuated far-UV luminosity. The arrows in Figures \ref{fig:ippe_psfuv}, \ref{fig:comparison_luv} and \ref{fig:ippe_beta} roughly show how corrections for dust would impact the production efficiencies ($\lesssim0.1$ dex lower). 

\item \cite{PrietoLyon_2023} studied a sample of $\sim100$ $z=3-7$ galaxies using HST and JWST/NIRCam photometry. The \ha\ flux was measured by comparing the observed photometry to the continuum flux, which was obtained from SED fitting using BAGPIPES, employing BC03 templates, an SMC dust law, and an exponentially rising delayed SFH. A wide range of values is obtained for the \ha\ flux, and this is reflected in wide range of \xiion\ values obtained.

\item \cite{Saxena_2023} studied 16 faint LAEs using spectroscopic data from the JWST Advanced Deep Extragalactic Survey (JADES). The \ha\ flux was used to estimate \xiion\ for all galaxies except one for which the \ha\ line was not within the spectral coverage. Dust was corrected using the \ha$/$\hb\ ratio (H$\gamma/$H$\beta$ for the aforementioned exception).
 
\item \cite{Schaerer_2022} measured the \oiii$\lambda$5007 EW of 3 $z\sim8$ galaxies in the SMACS field using JWST/NIRSpec. Using relations between \xiion\ and \oiii$\lambda$5007 at low-$z$, the authors obtained a rough estimate of $\log_{10}(\xi_{\rm ion}\rm{/erg^{-1}Hz})\approx25.1-25.5$. In Figure \ref{fig:ippe_psfuv}, we have plotted a representative data point at the midpoint of this estimate and the mean far-UV luminosity of the 3 galaxies.

\item \cite{Simmonds_2023} obtained both \xiion\ and \xidust\ for 30 $z\sim5.4-6.6$ galaxies, using the flux excess from JWST Extragalactic Medium-band Survey \citep[JEMS,][]{Williams_2023} observations to estimate the \ha\ flux. We note that most galaxies in the sample have low values of dust attenuation and hence the values of \xiion\ and \xidust\ are very similar in most cases, with a difference of at most $\sim0.1$ dex. In Figures \ref{fig:comparison_luv} and \ref{fig:ippe_beta}, we have plotted \xidust\ and indicated with an arrow the possible range of values, should dust-attenuation be accounted for.

\item \cite{Stark_2015} found evidence for the CIV$\lambda$1548 line in the KECK/MOSFire observation of a gravitationally-lensed $z=7.045$ galaxy (A1703-zd6). We note that the authors use the dust-attenuated far-UV luminosity to calculate the production efficiency, i.e. they measure \xidust. The estimate for \xidust\ was found by fitting the line emission and photometry to a grid of photoionisation models, generated with BC03 \citep{BC03} spectra.

\item \cite{Stark_2016} studied the Keck/MOSFire spectroscopy of 3 galaxies, chosen for their strong \oiii$+\rm{H}\beta$ emission. \xiion\ was obtained from a fit to emission line and photometric constraints, using \beagle, and assuming an exponentially delayed SFH superposed with a 10 Myr burst of constant star formation. In Figure \ref{fig:ippe_psfuv}, we plot the galaxy EGS-zs8-1, and in Figure \ref{fig:comparison_luv} we plot all 3 galaxies.

\item \cite{Stefanon_2022} stacked the SPITZER/IRAC photometry of $\sim100$ LBGs at $z\sim8$, and obtained \xiion\ by estimating the \ha\ flux from the photometry and assuming negligible dust attenuation.

\item \cite{Sun_2022} measured \xiion\ for 3 \ha\ $+$ \oiii$\lambda5007$ emitters, detected with JWST/NIRCam wide-field slitless spectroscopy (WFSS). \nion\ was obtained from the \ha\ line and the far-UV luminosity from SED fitting with \cigale, assuming an exponentially delayed SFH with an optional late starburst, and a Calzetti dust law.

\item \cite{Tang_2023} measured \xiion\ for 12 $z>7$ galaxies from their CEERS JWST/NIRSpec sample. \xiion\ is obtained by fitting to spectroscopic data and additional photometry using \beagle, assuming a CSFH and an SMC attenutation curve.

\item \cite{Whitler_2023} used JWST/NIRCam data to study a sample of 28 galaxies at $z\sim8.4-9.1$, in the vicinity of two luminous LAEs in the EGS field (the sample includes one of the luminous LAEs). \xiion\ was obtained by SED fitting with the \beagle\ tool, assuming a CSFH and an SMC dust law.
\end{itemize}

\subsection{Far-UV luminosity}


Figure \ref{fig:ippe_psfuv} shows predictions for the production efficiency alongside a number of observations. For the galaxy population in \flares\ at $z=5-10$ with $M_{\star}>10^8\;\rm{M_\odot}$ and above the dust-attenuated luminosity threshold $\log_{10}(L_{\rm FUV, dust}/\mathrm{erg\,s^{-1}Hz^{-1}})=28$, we obtain a median production efficiency of $\log_{10}(\xi_{\rm ion}\rm{/erg^{-1}Hz})=25.40^{+0.16}_{-0.17}$, and a 2$\sigma$ range of $\log_{10}(\xi_{\rm ion}\rm{/erg^{-1}Hz})=25-25.7$.

There is a strong positive correlation between the far-UV luminosity and stellar mass of galaxies in \flares, hence we expect the production efficiency to follow similar trends to the specific emissivity. The negative trend at high stellar masses in the lowermost panel of Figure \ref{fig:ippe_mstar_col} is mirrored in Figure \ref{fig:ippe_psfuv}, where we see that brighter galaxies tend to have lower production efficiencies. We attribute this trend to metallicity and age increasing with far-UV luminosity. The trend is flatter at low luminosities because the mass cut omits bright, lower-mass galaxies with dust-attenuated luminosities in the range $\log_{10}(L_{\rm FUV,\, dust}/\mathrm{erg\,s^{-1}Hz^{-1}})=28-28.5$.

On the whole, while there are some observations that sit within the range of values predicted by \flares\ \citep{Matthee_2022, Schaerer_2022, Lin_2023}, the observed values of \xiion\ cover a much wider range than those predicted in \flares, with a tendency towards higher values. For example, \cite{Endsley_2021} obtain \xiion\ values in the range $\log_{10}(\xi_{\rm ion}\rm{/erg^{-1}Hz})=25.4-26.2$, and \cite{Tang_2023} measure consistently high values, with $\log_{10}(\xi_{\rm ion}\rm{/erg^{-1}Hz})=25.6-26$. Other notably high measurements include those by \cite{Stefanon_2022}, \cite{Fujimoto_2023} and \cite{Endsley_2022}. We note the difference in representation of the datasets in Figure \ref{fig:ippe_psfuv} (and also throughout this work) -- data points for individual measurements have a transparent fill, while collections or stacks of galaxies have a solid fill, and may contain a wide spread of \xiion\ values that are not hinted at in the plot.

\cite{Bouwens_2016}, \cite{Faisst_2019} and \cite{Lam_2019} find within their samples a trend of higher production efficiencies at lower far-UV luminosities. Other observations of \xiion\ at high-redshift are not highly suggestive of such a trend with far-UV luminosity. It may be that sample bias makes it difficult to infer such a trend -- this is especially the case for studies targeting \oiii\ and Lyman-$\alpha$ emitters, which tend to be highly star-forming galaxies and hence have high production efficiencies. At low redshift, some studies have observed this trend confidently \citep[$3.8<z<5$ sample in][]{Bouwens_2016, Matthee_2017, Maseda_2020}, while others have not \citep{Shivaei_2018, Emami_2020}.

Theoretical predictions for \xiion\ tend to cover a similar range of values to \flares. \cite{Ceverino_2019} made predictions for \xiion\ with the \firstlight\ simulations, using v2.1 of the BPASS SPS library \citep{Eldridge_2017}. For galaxies in the approximate stellar mass range $10^6<M_*/\rm{M_\odot}<10^9$, they obtained values of \xiion\ between $\log_{10}(\xi_{\rm ion}\rm{/erg^{-1}Hz})\approx25-25.5$. In Figure \ref{fig:ippe_psfuv_mstar} of the Appendix, we show our results alongside those of \cite{Yung_2020}, both using the same BPASS library (binary v2.2.1), and find comparable values between the two, with \flares\ predicting a slightly higher median by $\sim0.1$ dex. 

There is no one reason that stands out as to why the discrepancy between theoretical and observed values of \xiion\ exists. In terms of observations, samples may be biased towards galaxies with the highest production efficiencies, as previously mentioned, or there may be uncertainties in the modelling of dust. For galaxies containing little dust (and this is often the case for highly-ionising galaxies such as LAEs), the dust attenuation law has a small impact on the production efficiency \citep{Harikane_2018, Bunker_2023}. At other times, however, the production efficiency is more sensitive to the dust curve used. For example, \cite{Bouwens_2016} find that in their galaxy sample, the SMC law gives slightly higher values of the production efficiency (by $\sim$0.1 dex) than the Calzetti law. \cite{DeBarros_2019} find that the Calzetti law leads to an increase in the production efficiency by $\sim0.2$ dex compared to the SMC law. On the theoretical side, there may be uncertainties with respect to the choice of SPS model \citep{Wilkins_2019, Jones_2022}. We showed in Section \ref{sec:theory} that the SPS model used has a large impact on \xiion. There are also limitations in the simulations that should be noted. With \flares\ specifically, we do not resolve lower mass galaxies ($M_*<10^8\,\rm{M_\odot}$) that may have higher production efficiencies. In Figure \ref{fig:ippe_psfuv_mstar}, the ionising photon production efficiency is plotted against stellar mass. Though there are fewer observations to compare with, some of the larger production efficiencies measured are associated with stellar masses below the mass resolution of \flares\ \citep{Endsley_2022, Fujimoto_2023, Tang_2023}. Future higher-resolution runs of \flares\ will allow us to push our study to lower masses. A final comment is that we have not accounted for the contribution from AGN, which would increase the total production efficiency of a galaxy. \cite{Simmonds_2023} find a galaxy with $\log_{10}(\xi_{\rm ion}\rm{/erg^{-1}Hz})=26.59$, which is likely to be an AGN. In \flares, the inclusion of AGN can push the total production efficiency of a galaxy to at most $\log_{10}(\xi_{\rm ion}\rm{/erg^{-1}Hz})\sim26$.

\subsection{Redshift dependence}
 
Figure \ref{fig:comparison_luv} shows that galaxies in \flares\ exhibit a weak trend of decreasing production efficiency with decreasing redshift, as a result of ageing stellar populations. This trend is observed for all luminosity bins. We note that \xiion\ has a stronger relation with far-UV luminosity than redshift -- values evolve between $0.05-0.1$ dex from $z=10$ to $z=5$ for a given luminosity bin, but the difference between the median \xiion\ of the highest and lowest luminosity bin at any one point is $\sim0.2$ dex. It is worth re-iterating here that the mass cut we have used omits bright, lower-mass galaxies with dust-attenuated luminosities in the range $\log_{10}(L_{\rm FUV,\, dust}/\mathrm{erg\,s^{-1}Hz^{-1}})=28-28.5$ -- thus the sample in the faintest bin (in green) is incomplete. This explains the similarity between the two faintest bins (in green and orange). Also plotted in grey are the median trends from \cite{Yung_2020}, predicting slightly lower production efficiencies than \flares, and a similarly weak trend with redshift. The observations shown in Figure \ref{fig:comparison_luv} tend to be higher than the $1\sigma$ range in \flares. This is likely due to bias towards high \xiion\ populations at high-redshift, with observations often focused on \oiii\ or Lyman-$\alpha$ emitters, as discussed earlier. 

\subsection{Observable properties}

\begin{figure*}
	\includegraphics[width=1.6\columnwidth]{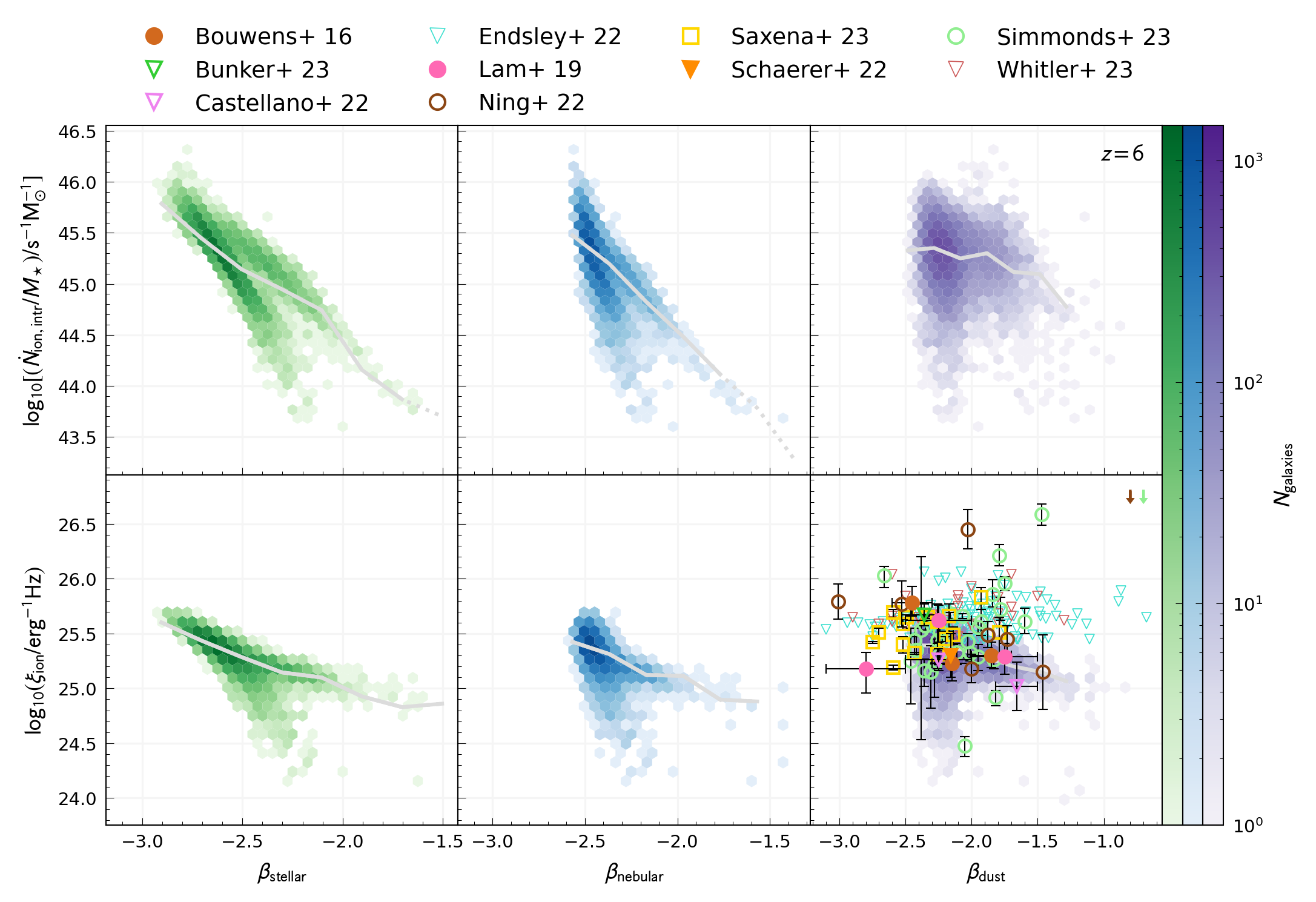}
	\caption{Specific emissivity (upper panels) and ionising photon production efficiency (lower panels) plotted against the UV-continuum slope, obtained from pure stellar SEDs (left column), stellar SEDs with nebular emission (middle column), stellar SEDs with nebular emission and dust (right column). Hex bins show the distribution of galaxies in \flares\ at $z=6$ and trend lines show the weighted median. Observations are represented by scatter points, with the same classification by marker style and fill as in Figure \ref{fig:ippe_psfuv}. Arrows indicate how values may change when accounting for dust \protect\citep[for measurements by][]{Ning_2022, Simmonds_2023}. To maintain readability, error bars have been left out for measurements by \protect\cite{Endsley_2022} and \protect\cite{Whitler_2023}.  \label{fig:ippe_beta}}
\end{figure*}

\begin{figure}
	\includegraphics[width=\columnwidth]{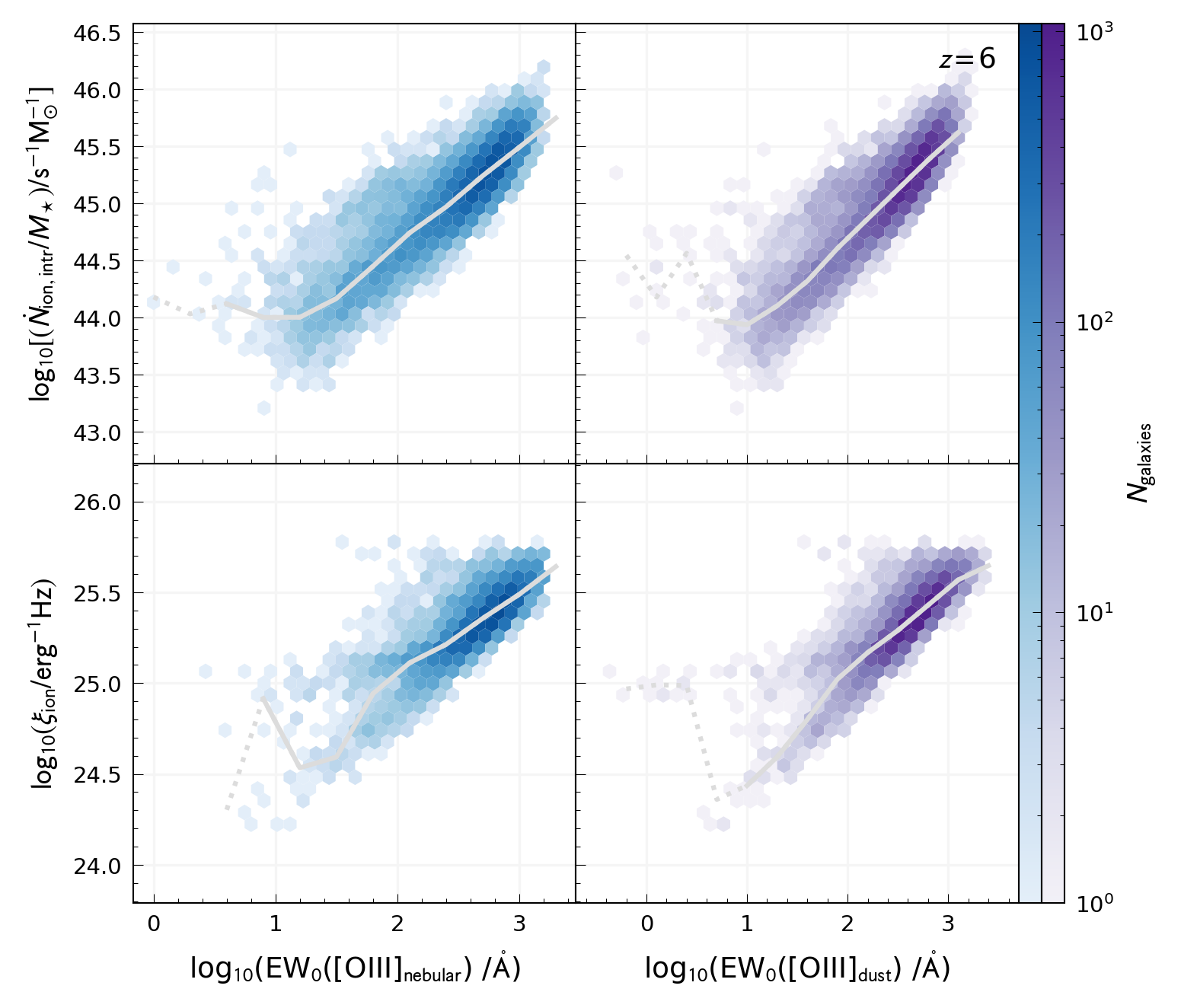}
	\caption{Specific emissivity (upper panels) and ionising photon production efficiency (lower panels) plotted against the intrinsic (left column) and dust-attenuated (right column) \oiii\ EW values. Hex bins show the distribution of galaxies in \flares\ at $z=6$ and trend lines show the weighted median. Observations are represented by scatter points. \label{fig:ippe_o3}}
\end{figure}

\subsubsection{UV-continuum slope}\label{sec:beta}

Figure \ref{fig:ippe_beta} shows how the specific emissivity and production efficiency vary with the UV-continuum slope, $\beta$. Moving from the leftmost column to the right shows the evolution of $\beta$ when nebular emission and dust are included in our SED modelling (thus the rightmost column shows the observed $\beta$). The general trend is that galaxies in \flares\ with high specific emissivities and production efficiencies tend to have bluer UV-continuum slopes. This trend is more evident when considering the pure stellar UV slopes, and subject to large amounts of scatter when considering the dust-attenuated values of $\beta$. We observe two distinct populations of galaxies, forming two branches in the distribution. The lower branch consists of low stellar mass galaxies with low metallicities, while the upper branch consists of more massive galaxies with higher metallicities. As we go through our analysis, Figure \ref{fig:stacked_beta} in the Appendix may be of interest to the reader, as it contains identical plots to Figure \ref{fig:ippe_beta} but coloured by specific star formation rate, metallicity, and stellar mass. 

To understand how the relation between the production efficiency and the observed value of $\beta$ (i.e. with nebular emission and dust) comes about, it is useful to first start with the UV slope derived from pure stellar SEDs. The lower left panel of Figure \ref{fig:ippe_beta} shows that galaxies with the highest production efficiencies have the bluest pure stellar UV slopes. This is as expected, since both phenomena are correlated with the presence of young, massive stars, and indeed we find that these galaxies have the highest specific star formation rates. The two `tails' we observe at redder values of $\beta$ are due to a difference in metallicity, as mentioned earlier -- the upper branch of galaxies has a higher metallicity, which leads to redder values of $\beta$. Due to the strong mass-metallicity relation in \flares, we find that galaxies in the upper branch are more massive as well. 

Moving to the lower middle panel, we see the evolution of $\beta$ when nebular emission is added. Note that we are only looking at the change in values of $\beta$, since the specific emissivity \nionM\ and the production efficiency \xiion\ remain the same throughout. The main difference is that the highly star-forming galaxies with high production efficiencies now have redder UV slopes, as a result of nebular continuum emission \citep{Byler_2017, Topping_2022}. 

Finally, moving on to the lower right panel, we see that the effect of dust is also to redden the UV slopes. The change is more pronounced for the upper branch of galaxies with high stellar masses and metallicities, as they tend to be more dusty. The addition of dust increases the amount of scatter in the relation -- in the case of the specific emissivity, this causes the median line to flatten considerably.

The predictions from \flares\ overlap with a number of observations \citep{Bouwens_2016, Castellano_2022, Schaerer_2022, Bunker_2023, Saxena_2023, Simmonds_2023}, but do not reproduce the elevated production efficiencies or the bluest values of $\beta$ ($\beta\sim-3.0$). The latter is due to the fact that all the ionising photons produced by a galaxy are reprocessed through the nebula, as described in Section \ref{sec:neb}. Previous studies have predicted an increase in \xiion\ for the bluest UV slopes \citep{Robertson_2013, Duncan_2015}. None of the observations plotted in Figure \ref{fig:ippe_beta} show a clear trend. This could be in part due to samples being biased towards highly star-forming galaxies, such as in the study by \cite{Endsley_2022}. More inclusive survey samples, or perhaps a greater variety of surveys, would enable us to come to a stronger conclusion.

\subsubsection{\oiii\ equivalent width}\label{sec:oiii}

Observational studies have found a positive correlation between \oiii\ emission line strength and the ionising photon production efficiency of galaxies \citep{Chevallard_2018, Reddy_2018, Tang_2019, Emami_2020, Castellano_2023}. This trend is also observed in \flares. We define the \oiii\ equivalent width (EW) as the combined equivalent widths of the \oiii\ doublet (\oiii$\lambda\lambda$4960,5008\AA). Figure \ref{fig:ippe_o3} shows that the specific emissivity and production efficiency of a galaxy is positively correlated with the \oiii\ EW, although the relation is subject to scatter. The positive correlation can be explained by the fact that \oiii\ emission is primarily driven by ionising radiation from young, massive stars, while the underlying optical continuum is boosted by emission from older stars as well. Thus the optical continuum can be interpreted as a `normalising' factor in the definition of the equivalent width. The distributions we observe are not strongly affected by dust -- this is something \cite{Wilkins_2023_OIII} also found in their paper analysing the \oiii\ emission of galaxies in \flares. We note that the gradient of the relation is metallicity-dependent. In fact, the spread of values seen in Figure \ref{fig:ippe_o3} consists of a low metallicity population and a high metallicity population, with different relations to the \oiii\ EW. More detail on this can be found in Section \ref{sec:app_o3} of the Appendix.

A solar abundance pattern was adopted to align with the BPASS SPS model that we use for our stellar SEDs. We note that high-redshift galaxies are $\alpha$-enhanced, meaning that they are likely to contain a higher proportion of $\alpha$-elements, such as oxygen \citep{Steidel_2016, Strom_2022, Cullen_2021, Byrne_2022}. \cite{Wilkins_2023_OIII} estimate a potential increase in the \oiii\ fluxes of $\sim0.1$ dex, when $\alpha$-enhancement is accounted for.

\subsection{Dust-attenuated production efficiency}\label{sec:dm1}

\begin{figure*}
	\includegraphics[width=1.8\columnwidth]{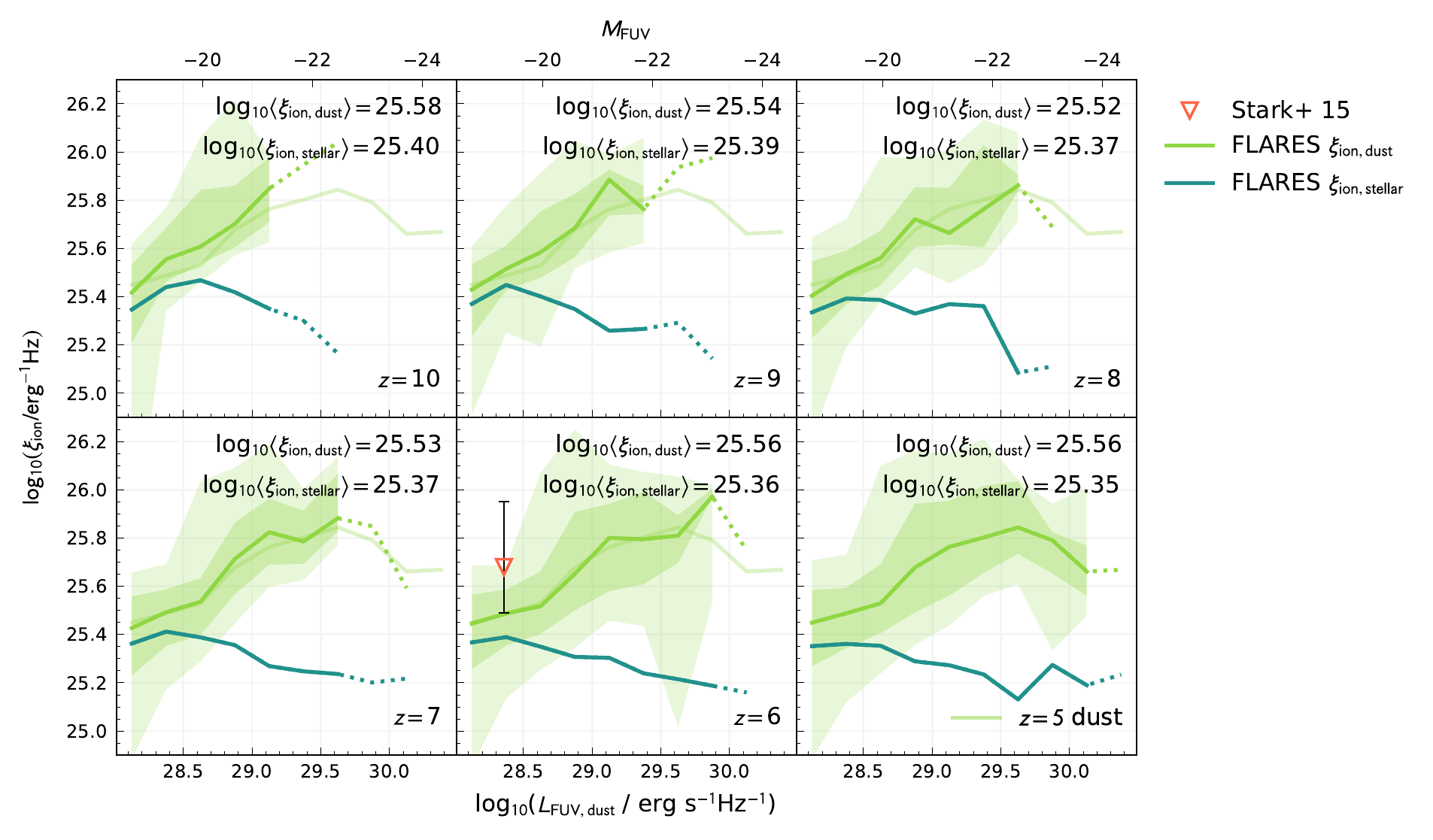}
	\caption{Ionising photon production efficiency of galaxies in \flares, defined in terms of dust-attenuated and pure stellar far-UV luminosity (\xidust\ and \xistellar\ in green and blue respectively). Trend lines show the weighted median production efficiencies, with dotted lines representing bins containing fewer than 10 galaxies. Shaded regions denote the 1 and $2\sigma$ range of \xidust. Observations are represented by scatter points.\label{fig:ippe_dm1fuv_compare}}
\end{figure*}

So far, we have defined the ionising photon production efficiency following Equation \ref{eq:xiion}, choosing to normalise the ionising emissivity of a galaxy by its intrinsic far-UV luminosity. From the perspective of forward-modelling from simulations, this means that no nebular emission or dust is modelled -- only stellar SEDs are used. For the sake of clarity, in this subsection we will label this value \xistellar, i.e. $\xi_{\rm ion}=\xi_{\rm ion,stellar}$. On the other hand, obtaining the intrinsic luminosity from observations requires an assumption of the dust model, which can be an uncertain parameter \citep[e.g.][]{Ferrara_2016, Behrens_2018, DeBarros_2019, Vijayan_2023}.

Alternatively, one can study the production efficiency defined using the dust-attenuated UV luminosity:
\begin{equation}
    \xi_{\rm ion,dust} = \frac{\dot{N}_{\rm ion, intr}}{L_{\rm UV,dust}},
\end{equation}
where the subscript `dust' is now used to clarify that the UV luminosity is dust-attenuated. Forward-modelling from simulations would thus require additional modelling assumptions, in the incorporation of nebular emission and dust, in order to obtain $L_{\rm UV,dust}$. On the other hand, working with \xidust\ can reduce the modelling assumptions made in observing the production efficiency, since the observed far-UV luminosity can be used without correcting for dust. This is more so the case when estimating the production efficiency from line fluxes, since an intrinsic SPS model has to be assumed to obtain $\dot{N}_{\rm ion, intr}$ from SED fitting. We note that using this definition does not completely remove the need for dust corrections -- estimating $\dot{N}_{\rm ion, intr}$ from line fluxes would require dust corrections in order to obtain the intrinsic flux. 

Figure \ref{fig:ippe_dm1fuv_compare} compares predictions for \xidust\ and \xistellar\ in \flares. We see no dependence of \xidust\ on redshift, with median values hovering around $\log_{10}(\xi_{\rm ion,dust}\rm{/erg^{-1}Hz})=25.55$. \xidust\ follows a reverse trend to \xistellar, increasing from faint luminosities to $\log_{10}(L_{\rm FUV, dust}\rm{/erg\,s^{-1}\,Hz^{-1}})\sim29.5$, after which values plateau or tentatively decrease. The ratio between \xidust\ and \xistellar\ roughly follows the trend in dust attenuation, by definition. Figure 9 of \cite{Vijayan_2021} shows the dust attenuation in the far-UV of galaxies in \flares, as a function of far-UV magnitude. Brighter, more massive galaxies tend to be more highly attenuated, as they have undergone longer periods of star formation that enrich the ISM and produce dust. However, the attenuation stops increasing at $\log_{10}(L_{\rm FUV, dust}\rm{/erg\,s^{-1}\,Hz^{-1}})\sim29.5$ ($M_{\rm FUV}\sim22$), which is also where we observe the downturn in \xidust.

%% file: sections/6.emissivity.tex
\begin{figure*}
	\includegraphics[width=1.6\columnwidth]{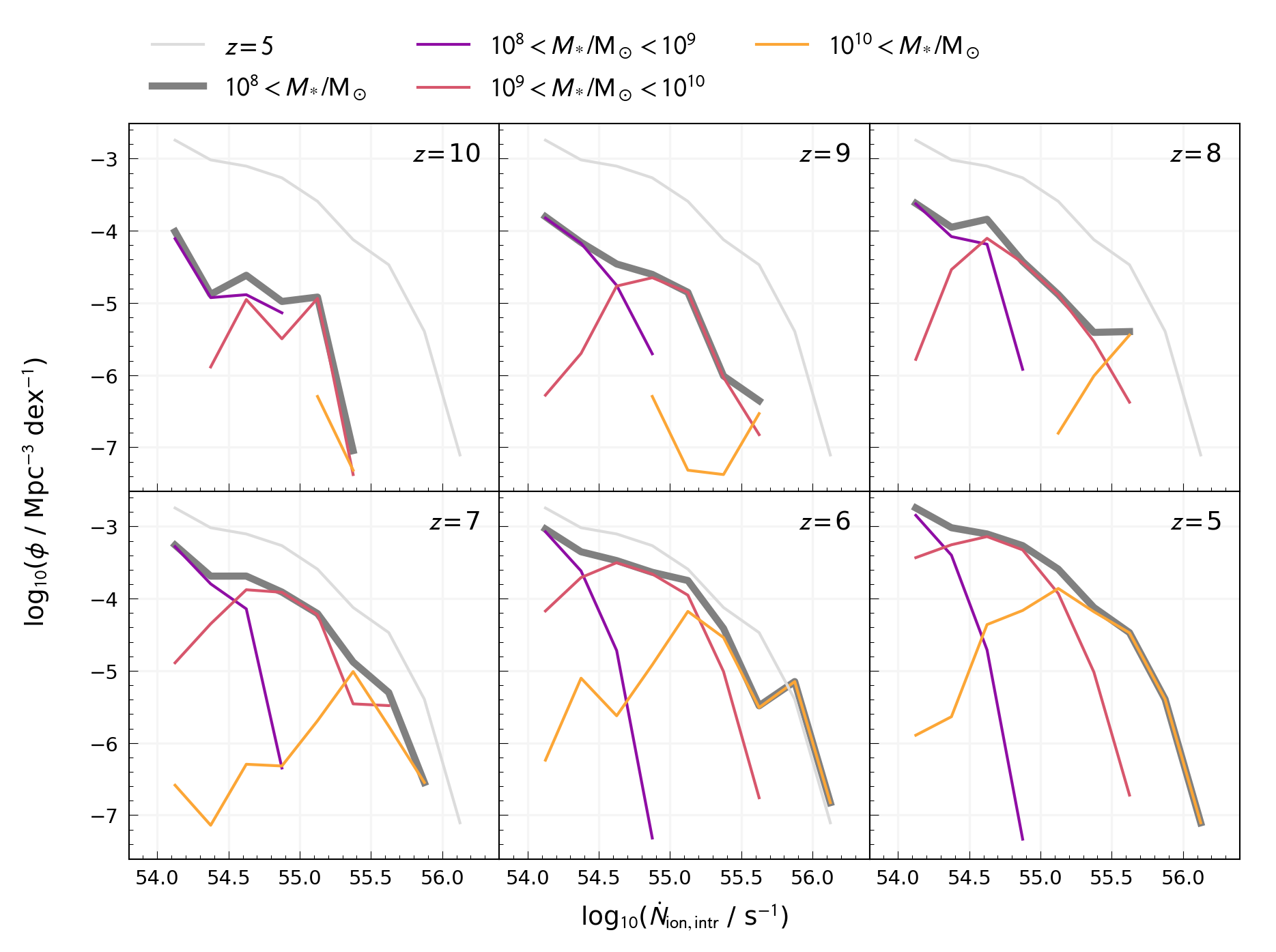}
	\caption{Lyman-continuum luminosity function (stellar contribution only) for galaxies in \flares. In dark grey is the luminosity function for all galaxies with ionising emissivity $\dot{N}_{\rm ion,intr}>10^{54}\rm{s^{-1}}$, which roughly corresponds to a stellar mass cut of $M_*>10^8\rm{M_\odot}$. This population of galaxies has been binned by stellar mass, and the contribution from each mass bin plotted as well (in order of increasing stellar mass: purple, red, yellow). As a reference, the $z=5$ function is plotted in light grey.\label{fig:n_ion}}
\end{figure*}

\begin{figure*}
	\includegraphics[width=1.6\columnwidth]{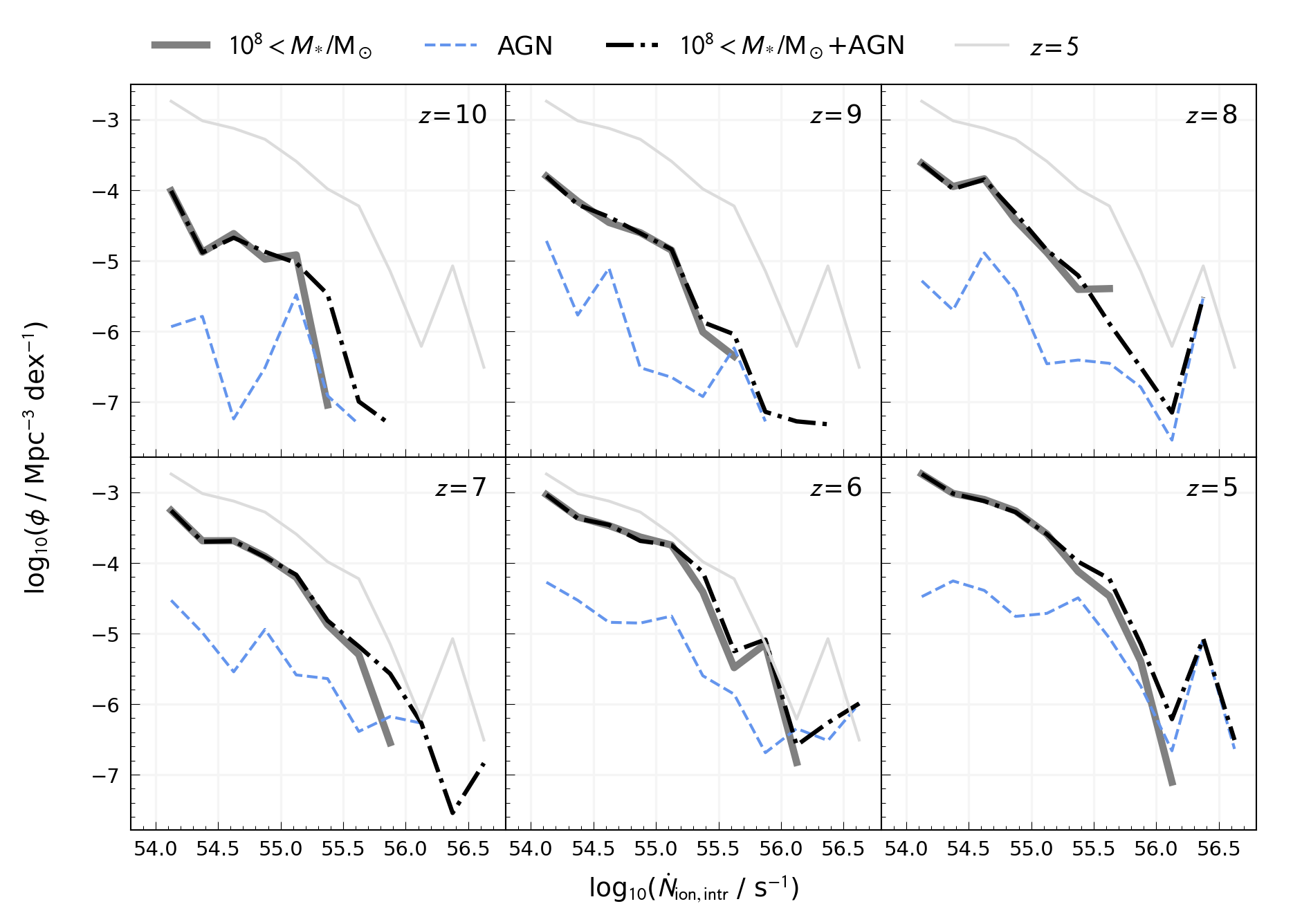}
	\caption{Lyman-continuum luminosity function for galaxies in \flares, considering all galaxies with stellar mass $M_*>10^8\rm{M_\odot}$. The solid, dark grey line shows the stellar luminosity function, the dashed blue line shows the AGN luminosity function, and the dash-dotted black line shows the luminosity function total luminosity (i.e. stellar $+$ AGN). As a reference, the $z=5$ function is plotted in light grey. \label{fig:n_ion_with_agn}}
\end{figure*}

\section{Intrinsic ionising emissivity}\label{sec:nion}

In this section, we study the intrinsic ionising emissivity of galaxies in \flares, analysing the intrinsic LyC luminosity function and the relative contributions to the total ionising emissivity from different populations in our sample.

\subsection{LyC luminosity function}\label{sec:lyclf}

Figure \ref{fig:n_ion} shows the evolution of the LyC luminosity function with redshift (note that only the contribution from stars is shown). The contributions from different stellar mass bins are also shown, with the higher mass bins dominating at the brighter end of the luminosity function. We showed in Sections \ref{sec:spec_nion} and \ref{sec:ippe} that massive galaxies tend to be less efficient at producing ionising radiation. Despite this, their large stellar populations compensate for the low efficiency, and we find that the total ionising emissivity of a galaxy increases with stellar mass. Hence, the LyC luminosity function is strongly governed by the galaxy stellar mass function -- the negative slopes observed in Figure \ref{fig:n_ion} are due to the decreasing number of galaxies as we go to higher stellar masses.

The AGN LyC luminosity function is shown as the dotted blue line in Figure \ref{fig:n_ion_with_agn} (we refer the reader to Kuusisto et al., in prep for information on how AGN emission is modelled in \flares). At lower emissivities, the AGN luminosity function is small compared to that of stars. However, AGN dominate the bright end of the LyC luminosity function -- after the rapid drop in the stellar luminosity function at $\rm{log}_{10}(\dot{N}_{\rm ion}/s^{-1})\sim56$, we find it is the more gently decreasing AGN luminosity function that extends the combined emissivity to higher values (by $\sim0.5$ dex).

\subsection{Contribution from different populations}

\begin{figure}
	\includegraphics[width=\columnwidth]{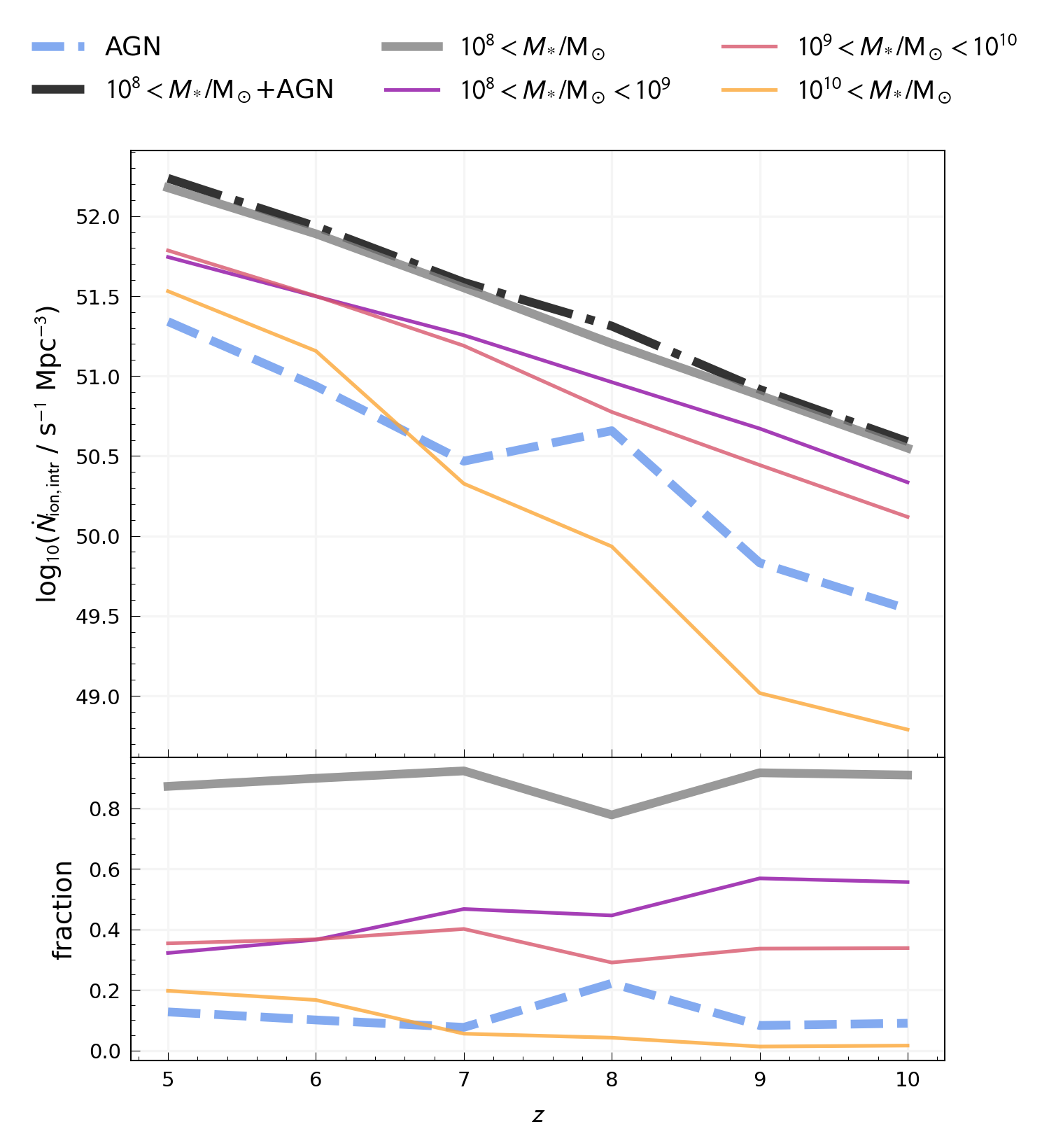}
	\caption{Top panel: redshift evolution of the total ionising emissivity per Mpc$^{-3}$ of \flares\ galaxies. Solid lines show the stellar contribution -- in dark grey is the contribution from galaxies with stellar mass $M_*>10^8\rm{M_\odot}$, in purple, red and yellow are the contributions from different stellar mass bins. The dashed blue line shows the AGN contribution, and the dash-dotted black line shows the total contribution (i.e. stellar $+$ AGN). Bottom panel: fraction of the total emissivity contributed by each population as a function of redshift. \label{fig:n_ion_count_comb}}
\end{figure}

Figure \ref{fig:n_ion_count_comb} shows the combined ionising emissivity per unit volume of different populations discussed in the previous section (\ref{sec:lyclf}), obtained by taking the integral of the LyC luminosity function. There is an increase in the total emissivity (black line) as redshift decreases, as one would expect, since the number of galaxies is increasing due to hierarchical assembly. Across all considered redshifts, from $z=10-5$, the contribution from AGN is small but still significant, generally providing between $10-20$\% of the total emissivity (lower panel of Figure \ref{fig:n_ion_count_comb}). The main source of ionising photons is lower mass galaxies -- from $z=10-7$, galaxies with $M_*=10^8-10^9\,\rm{M_\odot}$ contribute the most, with $M_*=10^9-10^{10}\,\rm{M_\odot}$ galaxies catching up at lower redshifts.

We note that these comparisons of the fractional contribution to the emissivity are relative statements that depend on the galaxy population being studied. If the analysis were extended to galaxies with stellar masses below the resolution in \flares, the fractional contribution from AGN would decrease, as AGN are less common in low mass galaxies. 

%% file: sections/conclusion.tex
\section{Conclusions}\label{sec:conclusion}

In this paper, we have used the \flare\ simulations to make predictions for the ionising properties of galaxies, namely, their intrinsic ionising emissivity and ionising photon production efficiency. We began by using simple toy models to explore how these two quantities vary with SFH, metallicity, choice of SPS model, and IMF. This provided a theoretical foundation for our main analysis on galaxies in \flares, which naturally have more realistic properties. We explored how the specific emissivity is linked to the physical properties of galaxies, and how the production efficiency relates to observable properties. We also compared our predictions to recent observational estimates. Our findings are summarised below:

\begin{itemize}

\item Using simple toy models, we show that the specific emissivity and production efficiency are sensitive to the SFH and metallicity of galaxies. In our examples, all SFHs cause a decline in both quantities. However, an increasing exponential SFH eventually leads to a plateau in their values, while an instantaneous SFH results in an immediate, sharp decline. Trends for a CSFH and decreasing exponential SFH fall between these two extremes. We also find that the specific emissivity and production efficiency strongly depend on the SPS library and IMF used.

\item The specific ionising emissivity of galaxies in \flares\ is strongly correlated with their specific star formation rate and negatively correlated with age, which we define as the initial mass-weighted median stellar age. This is because young, massive stars are the dominant source of ionising photons in a stellar population.

\item The specific ionising emissivity of galaxies in \flares\ shows little evolution at at low metallicities. We observe a peak in the specific emissivity at $Z_\star\sim10^{-2.5}$, after which there is a stronger trend of decreasing specific emissivity with increasing metallicity. The general trend can be attributed to the BPASS SPS model that we use, however, the peak is characteristic of galaxies in \flares\ -- we find that galaxies with stellar masses $M_\star\sim10^{9.5}\,\rm{M_\odot}$ (corresponding to a metallicity of $Z_\star\sim10^{-2.5}$) tend to undergo a burst of star formation in their cores, increasing the specific star formation rate and hence the specific emissivity.

\item Galaxies in \flares\ with stellar masses $M_\star\sim10^{9.5}\,\rm{M_\odot}$ exhibit a trend of decreasing specific emissivity with increasing stellar mass. This is due to the combined effects of increasing age and metallicity with increasing stellar mass, with metallicity likely playing a bigger role due to the weaker evolution of age with stellar mass.

\item The ionising photon production efficiency of galaxies in \flares\ generally increases as we go to lower far-UV luminosities. As the FUV luminosity of galaxies in \flares\ is strongly correlated with stellar mass, this trend parallels that of the specific emissivity with stellar mass, and likewise can be attributed to the effects of age and metallicity.

\item We find a trend of decreasing production efficiency as redshift decreases, due to the effect of increasing age. 

\item \flares\ predicts values of the production efficiency that are comparable with previous theoretical studies \citep{Wilkins_2016, Ceverino_2019, Yung_2020}. On the other hand, observations of the production efficiency tend to be around or higher than the values predicted by \flares. There are several possible reasons for this discrepancy. On the theoretical side, we are limited by the mass resolution of our model, and have not included the contribution from AGN. As for observations, we note that a variety of methods has been used to obtain the production efficiency. In general, measurements made using spectroscopically-obtained \ha\ or \hb\ line fluxes can be considered more robust than those obtained using SED fitting, or the line fluxes estimated from colours. We note that the choice of SPS model and dust model may have a strong influence on predicted values.

\item The production efficiency of galaxies in \flares\ generally decreases as we go to redder values of the UV continuum slope $\beta$, however this trend is subject to scatter and has a metallicity-dependent gradient. We observe a positive correlation between the production efficiency and the \oiii\ EW, although this relation is also subject to scatter.

\item Despite having lower specific ionising emissivities, it is the most massive galaxies that have the highest ionising emissivities, dominating the bright end of the intrinsic stellar Lyman-continuum luminosity function. We show that the luminosity function decreases at higher emissivities, governed by the galaxy stellar mass function. When including the AGN component, we find that AGN contribute relatively little compared to stars, but extend the luminosity function to higher emissivities.

\item Considering our galaxy population at $M_\star>10^8\,\rm{M_\odot}$, we find that the lowest mass galaxies contribute the most to the total ionising emissivity, due to their large population size. In general, the fractional contribution decreases with increasing stellar mass. 

\end{itemize}

The ionising emissivity and ionising photon production efficiency are important parameters in linking the formation and evolution of galaxies to the history of reionisation. In this paper, we have provided theoretical insight into these parameters, and made predictions based on the standard cosmological model. In large part thanks to the operations of JWST, the number of robust observational measurements of the production efficiency is steadily increasing, and will continue to do so in the coming years, enabling better constraints on the ionising properties of high-redshift galaxies.

%% file: sections/appendix.tex
\section{Ionising photon production efficiency}

\subsection{Stellar mass}

\begin{figure*}
	\includegraphics[width=2\columnwidth]{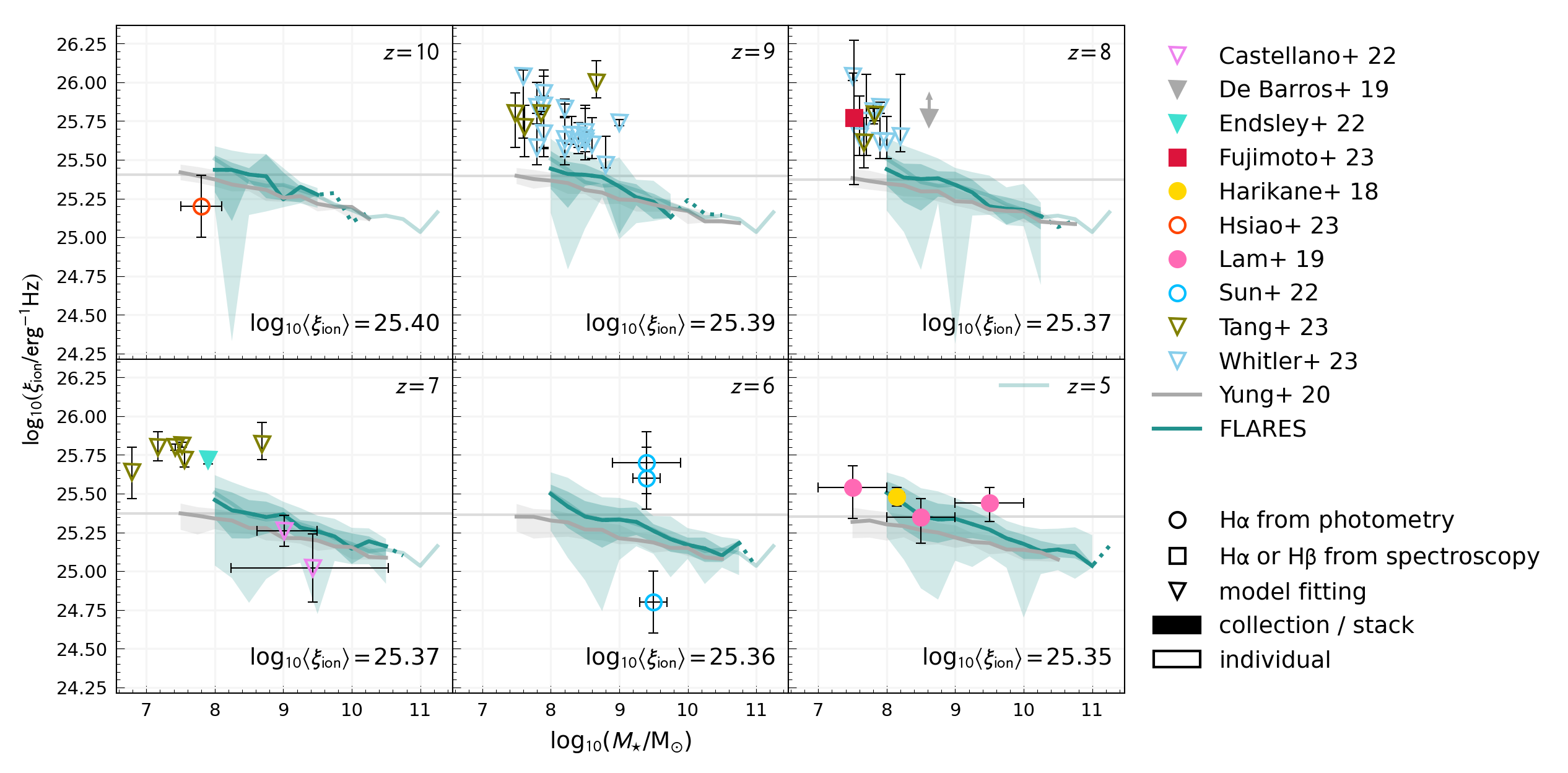}
	\caption{Ionising photon production efficiency as a function of stellar mass, for redshifts $5-10$. Trend lines and their colour-associated shaded regions represent the median and 1$\sigma$ range from theoretical models: in grey is the median trend from \protect\cite{Yung_2020}; in blue are the weighted median trends from \flares. The faint, horizontal grey line indicates the weighted mean of the \flares\ sample at each redshift. The translucent blue line plotted across all panels shows the \flares\ weighted median at $z=5$. Observations are displayed as scatter points: those with a transparent fill are measurements of individual galaxies; those with a solid fill are aggregated values, representing either stacks or collections of galaxies; circular and square data points represent measurements of the production efficiency obtained using Balmer emission line fluxes fluxes from photometry and spectroscopy respectively; triangular data points represent measurements of the production efficiency obtained from model fitting (this is a broad term that encompasses SED fitting). Observations are plotted in the panel corresponding to the nearest integer redshift. \label{fig:ippe_psfuv_mstar}}
\end{figure*}

Figure \ref{fig:ippe_psfuv_mstar} shows the production efficiency as a function of stellar mass for galaxies in \flares. The production efficiency generally decreases as stellar mass increases, similar to the trend seen in Figure \ref{fig:ippe_psfuv}. Also plotted are predictions from \cite{Yung_2020}, and measurements from several observational studies. Note that for the sake of readability, horizontal error bars have been omitted for measurements by \cite{Tang_2023} and \cite{Whitler_2023}. In general, \flares\ predicts slightly higher median values of the production efficiency than \cite{Yung_2020}, by $\sim0.1$ dex. Both \flares\ and \cite{Yung_2020} use v2.2.1 of the BPASS SPS library.

\subsection{UV-continuum slope}

\begin{figure}
	\includegraphics[width=\columnwidth]{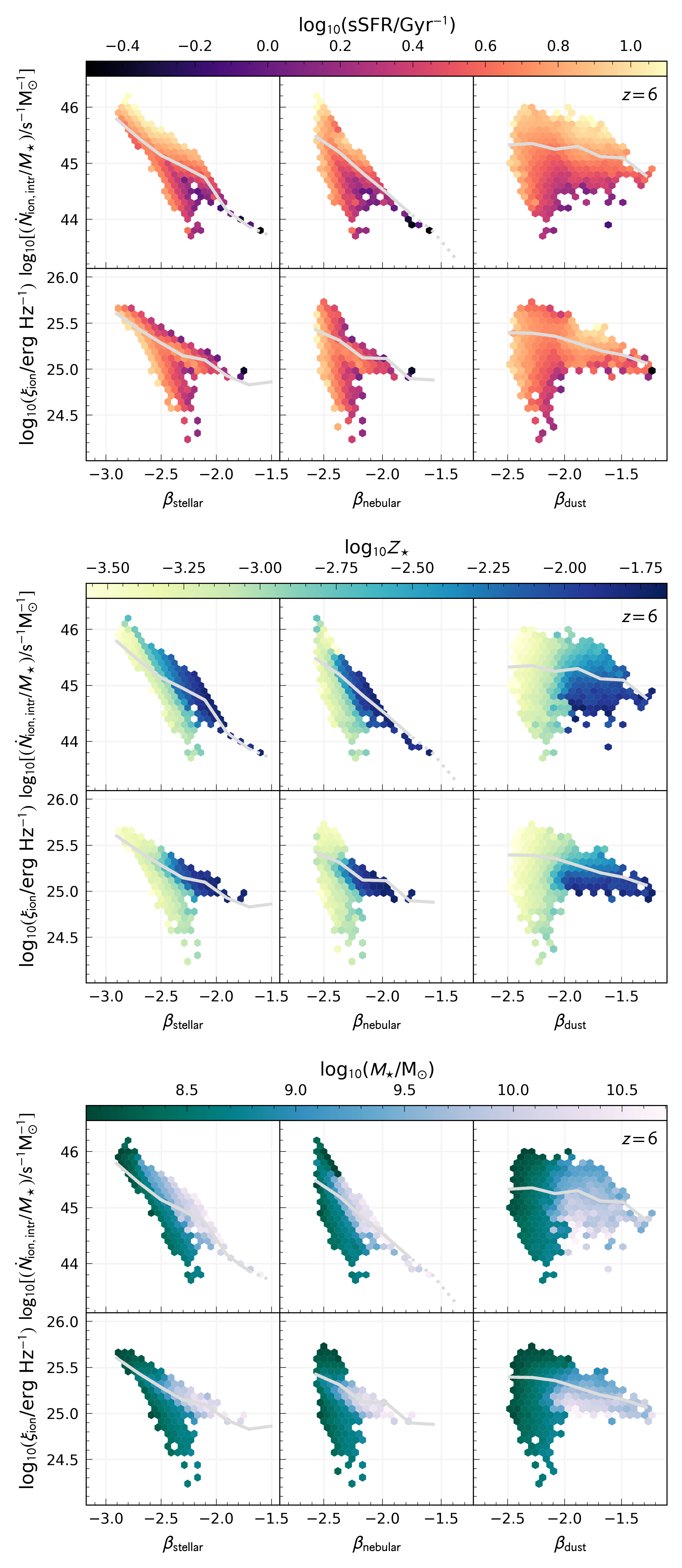}
	\caption{These three plots are identical except for the quantity used to colour the hex bins. Each plot shows the specific emissivity (upper panels) and ionising photon production efficiency (lower panels) plotted against the UV-continuum slope, obtained from pure stellar SEDs (left column), stellar SEDs with nebular emission (middle column), stellar SEDs with nebular emission and dust (right column). Hex bins show the distribution of galaxies in \flares\ at $z=6$, and are coloured by the mean specific star formation rate (top plot), stellar metallicity (middle plot), and stellar mass (bottom plot). Trend lines show the weighted median. \label{fig:stacked_beta}}
\end{figure}

Figure \ref{fig:stacked_beta} shows the specific emissivity and production efficiency as a function of the UV-continuum slope. This plot is supplementary to the discussion in Section \ref{sec:beta}.

\subsection{\oiii\ equivalent width}\label{sec:app_o3}

\begin{figure}
    \centering
	\includegraphics[width=0.95\columnwidth]{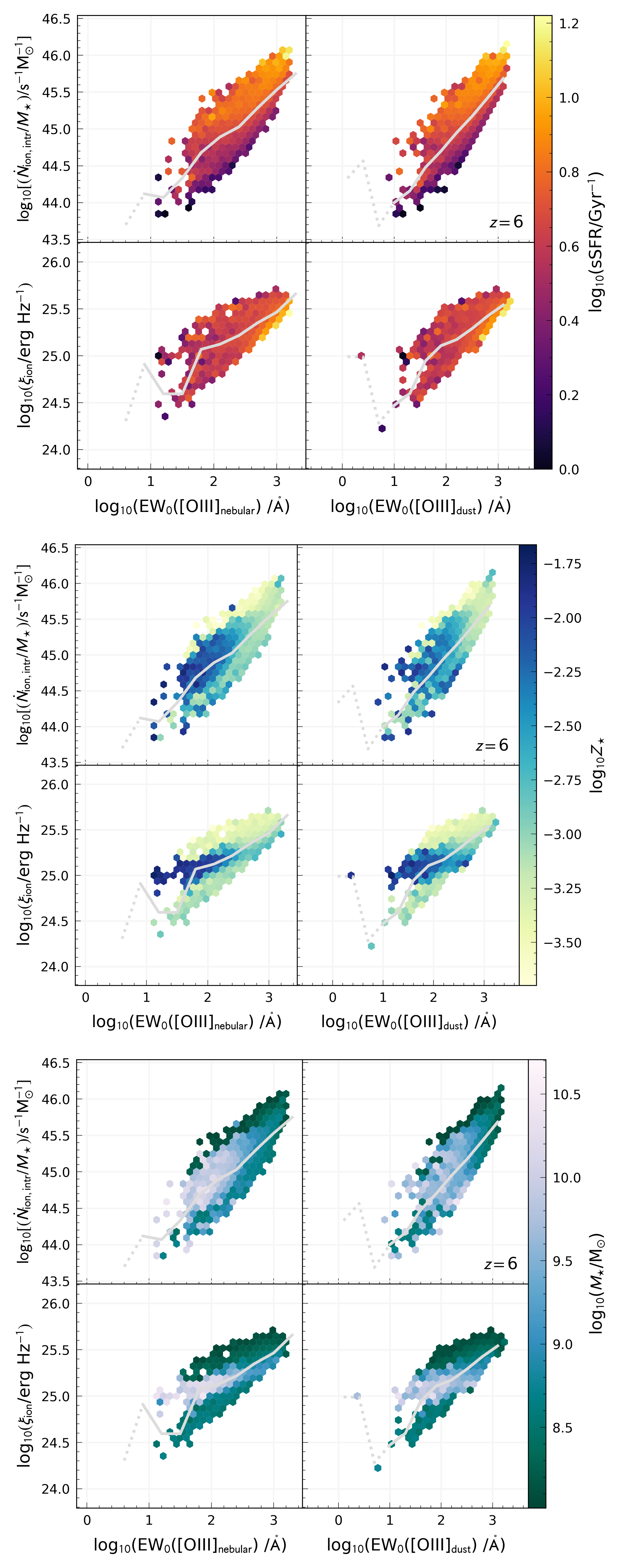}
	\caption{These three plots are identical except for the quantity used to colour the hex bins. Each plot shows the specific emissivity (upper panels) and ionising photon production efficiency (lower panels) plotted against the intrinsic (left column) and dust-attenuated (right column) \oiii\ EW values. Hex bins show the distribution of galaxies in \flares\ at $z=6$, and are coloured by the mean specific star formation rate (top plot), stellar metallicity (middle plot), and stellar mass (bottom plot). Trend lines show the weighted median. \label{fig:stacked_o3}}
\end{figure}

Two separate galaxy populations are observed in the relationship between the \oiii\ equivalent width (EW) and \xiion. This becomes clearer when looking at the middle and bottom plots in Figure \ref{fig:stacked_o3} -- there is a population of galaxies with higher stellar masses and metallicities that exhibits a weaker trend with \oiii\ EW. This is due to the effect of metallicity -- \cite{Wilkins_2023_OIII} show that the \oiii\ EW increases with metallicity (due to the increasing Oxygen abundance) until $Z_\star\sim10^{-2.5}$, and then decreases as a result of higher metallicities decreasing the amount of ionising radiation produced (second panel of Figure \ref{fig:ippe_mstar_col}). The population of galaxies exhibiting this weaker trend has metallicities above this critical value. We find that the two populations are not so distinct when considering the specific ionising emissivity.